\documentclass[journal,10pt,onecolumn]{IEEEtran}
\usepackage{cite}
\usepackage{color,yfonts}
\usepackage[pdftex]{graphicx}
\graphicspath{{fig/}{jpeg/}}
\usepackage[cmex10]{amsmath}
\usepackage{amssymb}
\usepackage{algorithm}
\usepackage{algpseudocode}
\usepackage{amsmath,amssymb,lipsum}
\usepackage{hyperref}
\usepackage{comment}
\usepackage{graphicx}
\usepackage[font=small]{caption}
\usepackage{subcaption}
\usepackage{mathtools,enumerate}
\input{mysymbol.sty}
\usepackage{needspace}

\usepackage{theorem}
\newtheorem{thm}{Theorem}
\newtheorem{lemma}{Lemma}
\newtheorem{proposition}{Proposition}
\newtheorem{corollary}{Corollary}

\theoremstyle{definition}
\newtheorem{assumption}{Assumption}
\newtheorem{remark}{Remark}

\def \QED {$\blacksquare$}

\newcommand{\qed}{\hfill\blacksquare}
 
\title{Proximity Without Consensus in Online \\ Multi-Agent Optimization}
\author{Alec Koppel$^\star$, Brian M. Sadler$^{\dagger}$, and Alejandro Ribeiro$^\star$
\thanks{Work in this paper is supported by NSF CCF-1017454, NSF CCF-0952867, ONR N00014-12-1-0997, ARL MAST CTA, and ASEE SMART. Part of the results in this paper appeared in \cite{c_Koppel2016a,c_Koppel2016b}.}
\thanks{ \noindent$^\star$Department of Electrical and Systems Engineering, University of Pennsylvania, 200 South 33rd Street, Philadelphia, PA 19104. Email: \{akoppel, aribeiro\}@seas.upenn.edu.}
\thanks{\noindent $^{\dagger}$U.S. Army Research Laboratory, Computation and Information Sciences Directorate, 2800 Powder Mill Road, Adelphi, MD 20783. Email: brian.m.sadler6.civ@mail.mil} 
}

\begin{document}
\maketitle

\begin{abstract}
We consider stochastic optimization problems in multi-agent settings, where a network of agents aims to learn parameters which are optimal in terms of a global objective, while giving preference to locally observed streaming information. To do so, we depart from the canonical decentralized optimization framework where agreement constraints are enforced, and instead formulate a problem where each agent minimizes a global objective while enforcing network proximity constraints. This formulation includes online consensus optimization as a special case, but allows for the more general hypothesis that there is data heterogeneity across the network. To solve this problem, we propose using a stochastic saddle point algorithm inspired by Arrow and Hurwicz. This method yields a decentralized algorithm for processing observations sequentially received at each node of the network. Using Lagrange multipliers to penalize the discrepancy between them, only neighboring nodes exchange model information.
We establish that under a constant step-size regime the time-average suboptimality and constraint violation are contained in a neighborhood whose radius vanishes with increasing number of iterations. As a consequence, we prove that the time-average primal vectors converge to the optimal objective while satisfying the network proximity constraints.
We apply this method to the problem of sequentially estimating a correlated random field in a sensor network, as well as an online source localization problem, both of which demonstrate the empirical validity of the aforementioned convergence results.
\end{abstract}

%!TEX root = root.tex
\section{Introduction}\label{sec:intro}

We consider online multi-agent optimization problems, where a group of interconnected agents aim to minimize a global objective $f= \sum_i f_i$ which may be written as a sum of local objectives $f_i$ available at different nodes $i$ of a network $\mathcal{G}=(V,\mathcal{E})$. The problem is online because information upon which the local objectives depend is sequentially and locally received by each agent. We consider the setting where agents aim to keep their decision variables {\it close} to one another but {\it not coincide} in order to minimize this global objective while giving preference to possibly distinct local signals. The motivation for this problem comes from the fact that consensus optimization methods implicitly operate on the hypothesis that the distribution of observations at each node is identical, which does not hold for a variety of problems in signal processing \cite{1458275} and robotics \cite{KoppelEtal16a}.

%discussion of ways to enforce consensus constraint
%history of consensus optimization
Distributed optimization has a rich history in networked control systems, \cite{bullo2009distributed,cao2013overview,4545274,jadbabaie2003coordination} wireless systems \cite{ribeiro2012optimal,5508319}, sensor networks \cite{1326696,Schizas2008}, and machine learning \cite{6879615,6483403}. 
Prior approaches to this problem require each agent to keep a local copy of the global decision variable, and approximately enforce an agreement constraint between the local copies at each iteration. To do so, various information mixing strategies have been proposed in which agents combine local gradient steps with a weighted average of their neighbors variables \cite{DBLP:journals/corr/abs-1112-2972,RamNedicVeeravalli, Yuan2013},  dual reformulations where each agent ascends in the dual domain \cite{1506308,Jakubiec2013}, and primal-dual methods which combine primal descent with dual ascent \cite{Arrow1958,Nedic2009c,c_Koppel2014a,Koppel2015a,c_Koppel2015b,KoppelEtal16a,Ling2014a}. 
Stochastic optimization, which allow for sequential processing of observations as they are received, are mostly based upon stochastic approximation \cite{Robbins1951,Tsitsiklis1986}, and have been successfully applied to extend multi-agent problems to the online domain \cite{c_Koppel2015b,RamNedicVeeravalli}.

%why agreement
In distributed optimization problems, agent agreement may not always be the primary goal. In large-scale settings where one aims to leverage parallel processing architectures to alleviate computational bottlenecks, agreement constraints are suitable.  In contrast, if there are different priors on information received at distinct subsets of agents, then requiring the network to reach a common decision may to degrade local predictive accuracy. Specifically, if the observations at each node are independent but \emph{not} identically distributed, consensus yields the \emph{wrong} solution. Moreover, there are tradeoffs in complexity and communications, and it may be that only a subset of nodes requires a solution.

In this paper, we seek to solve problems in which each agent aims to minimize a global cost $\sum_i f_i$ subject to a {network proximity constraint}, which allows agents the leeway to select actions which are good with respect to a global cost while not ignoring the structure of locally observed information. This setting may correspond to a multi-target tracking problem in a sensor network or a collaborative learning task in a robotic network where each robot is operating in a distinct domain. We design multi-agent optimization strategies where agents reach a common understanding of global information, while still retaining their local perspectives.

%paper outline
We propose a stochastic variant of the saddle point method \cite{Arrow1958,Nedic2009c} to solve online multi-agent optimization problems with network proximity constraints. Our main technical contribution is to demonstrate that saddle point iterates converge in expectation to a primal-dual optimal pair of this problem when a constant algorithm step-size is chosen. %Moreover, when a small constant step-size is chosen we establish the Lagrangian of the optimization problem linearly converges to a neighborhood of its saddle point.

We begin the paper in Section \ref{sec:prob} with a discussion of decentralized stochastic optimization problems with network proximity constraints and present an example to decentralized estimation of a correlated random field in a sensor network to illustrate key concepts. The saddle point algorithm is developed in Section \ref{sec:algorithm} by drawing parallels with deterministic and stochastic optimization, and exploiting factorization properties of the Lagrangian. The method relies on the definition of an augmented stochastic Lagrangian associated with the instantaneous global cost function, and operates by alternating primal descent and dual accept steps. 
In Section \ref{sec:convergence}, we establish the convergence properties of the proposed method to a primal-dual optimal pair in the constant step-size regime. We demonstrate the proposed method's practical utility on a spatially correlated random field estimation problem in a sensor network Section \ref{sec:fields}, and apply this tool to a decentralized online source localization problem in Section \ref{sec:sims} for a variety of problem instances. Finally, we conclude in Section \ref{sec:conclusion}.
%!TEX root = root.tex
\section{Problem Formulation}\label{sec:prob}

We consider agents $i$ of a symmetric and connected network $\ccalG = (V, \ccalE)$ with $|V|=N$ nodes and $|\ccalE|=M$ edges and denote as $n_i:=\{j:(i,j)\in\ccalE\}$ the neighborhood of agent $i$. Each of the agents is associated with a (non-strongly) convex loss function $f_i : \ccalX \times \Theta_i \rightarrow \reals$ that is parameterized by a decision variable $\bbx_i \in \ccalX\subset \reals^p$ and a random variable $\bbtheta_i \in \Theta_i$ with a proper distribution. Throughout, we assume $\ccalX$ is a compact convex subset of $\reals^p$ associated with the $p$-dimensional parameter vector of agent $i$. The functions $f_i(\bbx_i,\bbtheta_i)$ for different $\bbtheta_i$ are interpreted as observations of a statistical model with a possible goal for agent $i$ being the computation of the optimal local estimate, 
\begin{equation} \label{eq:dist_stoch_opt_local}
   \bbx_i^{\text{L}} := \argmin_{\bbx_i \in \ccalX} 
                        F_i(\bbx_i)
                     := \argmin_{\bbx_i\in \reals^p} 
                        \mbE_{\bbtheta_i}[f_i(\bbx_i,\bbtheta_i)]\; . 
\end{equation}
In the online settings considered here the functions $f_i(\bbx_i,\bbtheta_i)$ are termed instantaneous because they are observed at particular points in time associated with realizations of the random variable $\bbtheta_i$; see Section \ref{sec:algorithm}. We refer to $F_i(\bbx_i):=\mbE_{\bbtheta_i}[f_i(\bbx_i,\bbtheta_i)]$ as the local average function at node $i$.

When we consider the network as a whole we can define the stacked vector $\bbx=[\bbx_1,\ldots,\bbx_N]$, which is an element of the product set $\ccalX^N\subset \reals^{Np}$, and the aggregate function $F(\bbx) := \sum_{i=1}^N \mbE_{\bbtheta_i}[f_i(\bbx_i,\bbtheta_i)]$. It then follows that the set of problems in \eqref{eq:dist_stoch_opt} is equivalent to the aggregate problem 
\begin{equation} \label{eq:dist_stoch_opt}
    \bbx^{\text{L}} = \argmin_{\bbx\in \ccalX^N} F(\bbx) 
                   := \argmin_{\bbx \in \ccalX^N} \sum_{i=1}^N \mbE_{\bbtheta_i}[f_i(\bbx_i,\bbtheta_i)]\; . 
\end{equation}
For convenience, we further define the stacked instantaneous function as $f(\bbx, \bbtheta)=\sum_i f_i(\bbx_i, \bbtheta_i)$. That \eqref{eq:dist_stoch_opt_local} and \eqref{eq:dist_stoch_opt} describe the same problem is true because there is no coupling between the variables $\bbx_i$ at different agents. In many situations, however, the parameters $\bbx_i^{\text{L}}$ that different agents want to estimate are related. It then makes sense to couple decisions of different agents as a means of letting agents exploit each others' observations. Consensus optimization problems work on the hypothesis that all agents are interested in learning the same decision parameters $\bbx_i$ for all $i\in V$. In this case, we modify \eqref{eq:dist_stoch_opt} by introducing consensus constraints of the form
\begin{equation}\label{eq:consensus_constraint}
    \bbx_i = \bbx_j, \text{ for all } j\in n_i \; .
\end{equation}
For a connected network this constraint makes all variables $\bbx_i$ equal -- hence the definition as a consensus problem. This hypothesis implicitly only makes sense in cases where agents observe information drawn from a common distribution, which may be overly restrictive. In general, parameters of nearby nodes are expected to be close but are not necessarily all equal, as is the situation in, e.g., the estimation of a smooth field that is albeit not uniform. To model this situation we introduce a convex local proximity function with real-valued range of the form $h_{ij}(\bbx_i, \bbx_{j} )$ and a tolerance $\gamma_{ij}\geq 0$.  These are used to couple the decisions of agent $i$ to those of its neighbors $j \in {n_i}$ through the definition of the optimal estimates as the solution of the constrained optimization problem
\begin{alignat}{2} \label{eq:coop_stoch_opt}
\bbx^*:=\ 
&\argmin_{\bbx \in \reals^{Np} }\ &&\sum_{i=1}^N \mbE_{\bbtheta_i}[f_i(\bbx_i,\bbtheta_i)] \nonumber\\
 &\text{\ s.t. } \ &&{h_{ij}(\bbx_i, \bbx_{j} ) \leq \gamma_{ij}}, \quad\text{ for all }j\in n_i.
\end{alignat}
%
%\blue{The function $h_{ij}(\bbx_i, \bbx_{j} )$ depends on both $i$ and $j$. Therefore, we need to give it to subindices. This has to change throughout the paper.}
In the formulation in \eqref{eq:coop_stoch_opt} we assume that the proximity function $h_{ij}(\bbx_i, \bbx_{j} )$ that couples node $i$ to node $j$ is equivalent to the proximity function $h_{ji}(\bbx_j, \bbx_{i} )$ that couples node $j$ to node $i$, i.e., that for all $\bbx_i$ and $\bbx_{j}$ we have $h_{ij}(\bbx_i, \bbx_{j} ) = h_{ji}(\bbx_j, \bbx_{i} )$. This implies that the constraints $h_{ij}(\bbx_i, \bbx_{j} )\leq \gamma_{ij}$ and $h_{ji}(\bbx_j, \bbx_{i} )\leq \gamma_{ji}$ are redundant. We also define the stacked constraint function $h : \ccalX^N \rightarrow \reals^M$. We keep them separate to maintain symmetry of the algorithm derived in Section \ref{sec:algorithm}.

The consensus constraints in \eqref{eq:consensus_constraint} are a particular example of a proximity function $h_{ij}(\bbx_i, \bbx_{j} )$ but so is the norm constraint $\|\bbx_i - \bbx_j \|^2 \leq \gamma_{ij}$. This latter choice makes the estimates $\bbx_i^*$ and  $\bbx_j^*$ of neighboring nodes close to each other but not equal. Implicitly, this allows $i$ to incorporate the (relevant) information of neighboring nodes without the detrimental effect of incorporating the information of far away nodes that is only weakly correlated with the estimator of node $i$.

The goal of this paper is to develop an algorithm to solve \eqref{eq:coop_stoch_opt} in distributed online settings where nodes don't know the distribution of the random variable $\bbtheta_i$ but observe local instantaneous functions $f_i(\bbx_i,\bbtheta_i)$ sequentially. An important observation here is that the workhorse distributed gradient descent \cite{DBLP:journals/corr/abs-1112-2972,RamNedicVeeravalli, Yuan2013} and  dual decomposition methods \cite{1506308,Jakubiec2013} can't be used to solve \eqref{eq:coop_stoch_opt} because they work only when the constraints $h_{ij}(\bbx_i, \bbx_{j})$ are linear. We will see that a stochastic saddle point method can be distributed when the functions  $h_{ij}(\bbx_i, \bbx_{j})$ are not necessarily linear and converges to the solution of \eqref{eq:coop_stoch_opt} when local instantaneous functions $f_i(\bbx_i,\bbtheta_i)$ are independently sampled over time. Before developing this algorithm, we discuss a representative example to clarify ideas.

%%%%%%%%%%%%%%%%%%%%%%%%%%%%%%%%%%%%%%%%%%%%%%%%%%%%%%%%%%%%%%%%%%%%%%%%%%%
%%%   E   X   A   M   P   L   E   %%%%%%%%%%%%%%%%%%%%%%%%%%%%%%%%%%%%%%%%%
%%%%%%%%%%%%%%%%%%%%%%%%%%%%%%%%%%%%%%%%%%%%%%%%%%%%%%%%%%%%%%%%%%%%%%%%%%%
%
\medskip \noindent{\bf Example (LMMSE Estimation of a Random Field).}
A Gauss-Markov random field is one in which the value of the field at the location of sensor $i$, denoted by $\bbx_i$, is of interest. Consider a sequential estimation problem in which the nodes of the sensor network acquire noisy linear transformations of the field's value at their respective positions. Formally, let $\bbtheta_{i,t}\in\reals^q$ be the observation collected by sensor $i$ at time $t$. Observations $\bbtheta_{i,t}$ are noisy linear transformations $\bbtheta_{i,t}=\bbH_{i}\bbx_i+ \bbw_{i,t}$ of a signal $\bbx_i\in\reals^p$ contaminated with Gaussian noise $\bbw_{i,t} \sim \ccalN(0, \sigma^2 \bbI)$ independently distributed across nodes and time. Ignoring neighboring observations, the minimum mean square error local estimation problem at node $i$ can then be written in the form of \eqref{eq:dist_stoch_opt_local} with $f_i(\bbx_i,\bbtheta_i) =  \|\bbH_{i}\bbx_{i} - \bbtheta_{i}\|^2$. The quality of these estimates can be improved using the correlated information of adjacent nodes but would be hurt by trying to make estimates uniformly equal across the network. This problem specification can be captured by the mathematical formulation
\begin{alignat}{2} \label{eq:dist_rls}
\bbx^*:=\ 
&\argmin_{\bbx\in\ccalX^{N} }\ &&\sum_{i=1}^N \mbE_{\bbtheta_i}\Big[\|\bbH_{i}\bbx_{i} - \bbtheta_{i}\|^2\Big] \\
 &\text{\ s.t. } \ &&(1/2)\|\bbx_{i}- \bbx_{j} \|^2 \leq \gamma_{ij}, \quad\text{ for all }j\in n_i. \nonumber
\end{alignat}
The constraint $(1/2)\|\bbx_{i}- \bbx_{j} \|^2 \leq \gamma_{ij}$ makes the estimate $\bbx^*_{i}$ of node $i$ close to the estimates $\bbx^*_{j}$ of neighboring nodes $j\in n_i$ but not so close to the estimates $\bbx^*_{k}$ of nonadjacent nodes $k\notin n_i$. The problem formulation in \eqref{eq:dist_rls} is a particular case of \eqref{eq:coop_stoch_opt} with $f_i(\bbx_i,\bbtheta_i) = \|\bbH_{i}\bbx_{i} - \bbtheta_{i}\|^2$ and $h_{ij}(\bbx_i,\bbx_j )= (1/2)\|\bbx_{i}- \bbx_{j} \|^2$.

%%%%%%%%%%%%%%%%%%%%%%%%%%%%%%%%%%%%%%%%%%%%%%%%%%%%%%%%%%%%%%%%%%%%%%%%%%%
%%%   S   E   C   T   I   O   N   %%%%%%%%%%%%%%%%%%%%%%%%%%%%%%%%%%%%%%%%%
%%%%%%%%%%%%%%%%%%%%%%%%%%%%%%%%%%%%%%%%%%%%%%%%%%%%%%%%%%%%%%%%%%%%%%%%%%%
%
\section{Algorithm Development}\label{sec:algorithm}

Recall that a decentralized algorithm is one in which node $i$ has access to local functions $f_i(\bbx_i,\bbtheta_i)$ and local constraints $h_{ij}(\bbx_i, \bbx_{j} ) \leq \gamma_{ij}$ and exchanges information with neighbors $j\in n_i$ only. Recall also that the algorithm is further said to be online if the distribution of $\theta_i$ is unknown and agent $i$ has access to independent observations $\bbtheta_{i,t}$ that are acquired sequentially. Our goal is to develop an online decentralized algorithm to solve \eqref{eq:coop_stoch_opt}. To achieve this we consider the approximate Lagrangian relaxation of \eqref{eq:coop_stoch_opt} which we state as
\begin{align} \label{eq:lagrangian}
\ccalL(\bbx,\bblambda)&= \sum_{i=1}^N \Bigg[
 \mbE_{\bbtheta_i}[f_i(\bbx_i,\bbtheta_i)] \\
&\qquad +\frac{1}{2}\sum_{ j\in n_i} \left(
\lambda_{ij} \left(h_{ij}(\bbx_i,\bbx_j )\! -\! \gamma_{ij} \right)
 -\frac{\delta \eps_t}{2} \lambda_{ij}^2\right) \Bigg],\nonumber
\end{align}
where {$\lambda_{ij}\in  \reals^+$} 
 is a nonnegative Lagrange multiplier associated with the proximity constraint between node $i$ and node $j$. Observe that \eqref{eq:lagrangian} \emph{does not} define the Lagrangian of the optimization problem \eqref{eq:coop_stoch_opt}, but instead defines an \emph{augmented Lagrangian} due to the presence of the last term on the right-hand side. This last term $-(\delta \eps_t/2)\lambda_{ij}^2$, with scalar parameters $\delta$ and $\eps_t$, is a regularizer on the dual variable, whose utility arises in controlling the accumulation of constraint violation of the algorithm over time. See Section \ref{sec:convergence} for details. % us to control the magnitude of the primal gradient of the Lagrangian. %Specifically, the primal gradient of the augmented Lagrangian is proportional to the Lagrange multipliers, i.e. $\nabla_{\bbx}\ccalL(\bbx,\bblambda) \propto \sum_{ij} \lambda_{ij} \nabla_{\bbx} h_{ij}(\bbx_i, \bbx_j) $, and may become arbitrarily large for large $\lambda_{ij}$. The dual regularizer allows us to mitigate this issue. An alternative approach would be to restrict $\lambda_{ij}$ to a compact set, but we have found the regularization approach to be more analytically tractable.
 
 We propose applying a stochastic saddle point algorithm to \eqref{eq:lagrangian} which operates by alternating primal and dual stochastic gradient descent and ascent steps respectively. To do so, consider the stochastic approximation of the augmented Lagrangian evaluated at observed realizations $\bbtheta_{i,t}$ of the random variables $\bbtheta_{i}$, which we define as
\begin{align} \label{eq:stoch_lagrangian}
\hat{\ccalL}_t(\bbx,\bblambda)&= \sum_{i=1}^N\Big[ f_i(\bbx_i,\bbtheta_{i,t}) 
 + \frac{1}{2}\sum_{j\in n_i}
\lambda_{ij} \left(h_{ij}(\bbx_i,\bbx_j ) - \gamma_{ij} \right)
- \frac{\delta \eps_t}{2} \lambda_{ij}^2 \Big] . 
\end{align}
Define the stacked dual variable as $\bblam : = [\lam_1 ; \cdots ; \lam_M ] \in \reals^{M}$. Moreover, denote the network aggregate random variable as $\bbtheta = [{\bbtheta}_1 ; \cdots ; {\bbtheta}_N]$. Particularized to the stochastic Lagrangian stated in \eqref{eq:stoch_lagrangian}, the stochastic saddle point method takes the form 
\begin{align} \label{eq:sp_primal}
   \bbx_{t+1}   &=\ccalP_{\ccalX^N}\Big[ \bbx_t - \eps_t \nabla_\bbx \hat{\ccalL}_t (\bbx_t, \bblam_t)\Big] \;,  \\
      \bblam_{t+1} &= \Big[\bblam_t + \eps_t \nabla_{\bblam} \hat{\ccalL}_t (\bbx_{t}, \bblam_t)  \Big]_+\; ,
      \label{eq:sp_dual}
\end{align}
where  $\nabla_\bbx \hat{\ccalL} (\bbx_t, \bblam_t)$ and $\nabla_{\bblam} \hat{\ccalL} (\bbx_t, \bblam_t)$, are the primal and dual stochastic gradients of the augmented Lagrangian with respect to $\bbx$ and $\bblam$, respectively. These stochastic subgradients are approximations of the gradients of \eqref{eq:lagrangian} evaluated at the current realization of the random variable $\bbtheta$. The notation $\ccalP_\ccalX^N(\bbx)$ denotes component-wise orthogonal projection of the individual primal variables $\bbx_i$ onto the given convex compact set $\ccalX$, and $[ \cdot ]_+$ denotes the projection onto the $M$-dimensional nonnegative orthant $\reals^M_{+}$. As an abuse of notation, we also use $[ \cdot ]_+$ to denote scalar positive projection where appropriate. 

The method stated in \eqref{eq:sp_primal} - \eqref{eq:sp_dual} can be implemented with decentralized computations across the network, as we state in the following proposition.

%%%%%%%%%%%%%%%%%%%%%%%%%%%%%%%%%%%%%%%%%%%%%%%%%%%%%%%%%%%%%%%
%%%   P   R   O   P   O   S   I   T   I   O   N   %%%%%%%%%%%%%
%%%%%%%%%%%%%%%%%%%%%%%%%%%%%%%%%%%%%%%%%%%%%%%%%%%%%%%%%%%%%%%
%
\begin{proposition}\label{prop1} %\red{Tell me what the $\tilde{}$ variables are.}
Let $\bbx_{i,t}$ be the $i$th component of the primal iterate $\bbx_t$ and $\lam_{ij,t}$ the $i,j$th component the dual iterate $\bblam_t$. The primal variable update is equivalent to the set of $N$ parallel local variable updates
 \begin{align}\label{eq:local_primal_update}
    \bbx_{i,t+1} 
       = \ccalP_{\ccalX}\Big[   \bbx_{i,t}  - \eps_t  
          \Big(   \nabla_{\bbx_i} f_i (\bbx_{i,t} ; \bbtheta_{i,t} )
	   + \sum_{j \in n_i} (\lam_{ij,t}+\lam_{ji,t}) \nabla_{\bbx_i} h_{ij}  (\bbx_{i,t}, \bbx_{j,t})  \Big)\Big]\;.
\end{align}
Likewise, the dual variable updates in \eqref{eq:sp_dual} are equivalent to the $M$ parallel updates 
\begin{equation} \label{eq:local_dual_update}
   \lam_{ij,t+1} \!=  \! \Big[\! (1\! - \eps_t \delta) \lam_{ij,t} + \eps_t \left( h_{ij}  (\bbx_{i,t}, \bbx_{j,t})\! - \!\gamma_{ij} \!\right)\Big]_+  .
\end{equation}\end{proposition}

%%%%%%%%%%%%%%%%%%%%%%%%%%%%%%%%%%%%%%%%%%%%%%%%%%%%%%%%%%%%%%%
%%%   P   R   O   O   F   %%%%%%%%%%%%%%%%%%%%%%%%%%%%%%%%%%%%%
%%%%%%%%%%%%%%%%%%%%%%%%%%%%%%%%%%%%%%%%%%%%%%%%%%%%%%%%%%%%%%%
%
\noindent{\bf Proof: } See Appendix A. $\qed$ \medskip
%%%%%%%%%%%%%%%%%%%%%%%%%%%%%%%%%%%%%%%%%%%%%%%%%%%%%%%%%%%%%%%%%%
%%%   A   L   G   O   R   I   T   H   M   %%%%%%%%%%%%%%%%%%%%%%%%
%%%%%%%%%%%%%%%%%%%%%%%%%%%%%%%%%%%%%%%%%%%%%%%%%%%%%%%%%%%%%%%%%%
%
{\small \begin{algorithm}[t] 
\caption{SSPM: Stochastic Saddle Point Method}
{\small \label{alg:sspm} 
\begin{algorithmic}[1]
\Require initialization $\bbx_{0}$ and $\bblam_{0}=\bb0$, step-size $\eps_t$, regularizer $\delta$
\For {$t=1,2,\ldots, T$}
\Loop {{\bf{ in parallel}} agent $i\in V$}
      \State Send primal and dual variables $\bbx_{i,t}, \bblam_{ij,t}$ to nbhd. $j\in n_i$
      \State Receive variables $\bbx_{j,t}, \bblam_{ij,t}$ from neighbors $j\in n_i$
  %   \State Send primal variable $\bbx_{j,t}$ to neighbors $ k\in n_j$, receive $\bbx_{k,t}$
   \State Update local parameter $\bbx_{i,t}$ with \eqref{eq:local_primal_update}
          \begin{align*}
    \bbx_{i,t+1} 
       = \ccalP_{\ccalX}\Big[  & \bbx_{i,t}  - \eps_t  
          \Big(   \nabla_{\bbx_i} f_i (\bbx_{i,t} ; \bbtheta_{i,t} )
	   + \sum_{j \in n_i} (\lam_{ij,t}+\lam_{ji,t}) \nabla_{\bbx_i} h_{ij}  (\bbx_{i,t}, \bbx_{j,t})  \Big)\Big]\;.
\end{align*}
\EndLoop
\Loop {{\bf{ in parallel}} communication link $(i,j)\in \ccalE$}
   \State Update dual variables at network link $(i,j)$ [cf. \eqref{eq:local_dual_update}]
\begin{align*} 
\lam_{ij,t+1} \!=  \! \Big[\! (1\! - \eps_t \delta) \lam_{ij,t} + \eps_t \left( h_{ij}  (\bbx_{i,t}, \bbx_{j,t})\! - \!\gamma_{ij} \!\right)\Big]_+
\end{align*}
\EndLoop
\EndFor
\end{algorithmic}}
\end{algorithm}}
%%%%%%%%%%%%%%%%%%%%%%%%%%%%%%%%%%%%%%%%%%%%%%%%%%%%%%%%%%%%%%%
%%%   M   A   I   N       M   A   T   T   E   R   %%%%%%%%%%%%%
%%%%%%%%%%%%%%%%%%%%%%%%%%%%%%%%%%%%%%%%%%%%%%%%%%%%%%%%%%%%%%%
%

With primal variables $\bbx_{i,t}$ and Lagrange multipliers $\lam_{ij,t}$ maintained and updated by node $i$, Proposition \ref{prop1} implies that the saddle point method in \eqref{eq:sp_primal}-\eqref{eq:sp_dual} can be translated into a decentralized protocol in which: (i) The primal and dual variables variables of distinct agents across the network are decoupled from one another. (ii) The updates require exchanges of information among neighboring nodes only. This protocol is summarized in Algorithm \ref{alg:sspm}.

Indeed, in the primal update in \eqref{eq:local_dual_update} agent $i$ can compute the stochastic gradient $\nabla_{\bbx_i} f_i (\bbx_{i,t} ; \bbtheta_{i,t} )$ of its objective function by making use of its local observations $\bbtheta_{i,t}$ and its decision variable $\bbx_{i,t}$ at the previous time slot $t$. To compute the gradients of the constraint functions $\nabla_{\bbx_i} h_{ij}  (\bbx_{i,t}, \bbx_{j,t})$ the primal variables $\bbx_{j,t}$ of neighboring nodes $j\in n_i$ are needed on top of the local variables $\bbx_{i,t}$, but these can be communicated from neighbors. To implement \eqref{eq:local_primal_update} agent $i$ also needs access to the Lagrange multipliers $\lam_{ij,t}$ associated with the network proximity constraints $h_{ij}  (\bbx_{i}, \bbx_{j})$ and the multipliers $\lam_{ji,t}$ associated with the network proximity constraints $h_{ji}(\bbx_{j}, \bbx_{i})$. The multipliers $\lam_{ij,t}$ are locally available at $i$ and the multipliers $\lam_{ji,t}$ can be communicated from neighbors.

To implement the dual update in \eqref{eq:local_dual_update} agent $i$ needs access to its own dual variable $\lam_{ij,t}$ as well as the local decision variables $\bbx_{i,t}$. It also needs access to the primal variables $\bbx_{j,t}$ of neighbors $j\in n_i$ to compute the local dual gradient which is given as the constraint slack $h_{ij}  (\bbx_{i,t}, \bbx_{j,t}) - \gamma_{ij}$. As in the primal, these neighboring variables can be communicated from neighbors. We can then implement \eqref{eq:local_primal_update} after nodes exchange primal and dual variables $\bbx_{i,t}$ and $\lam_{ij,t}$, proceed to implement \eqref{eq:local_dual_update} after they exchange updated primal variables $\bbx_{i,t}$, and conclude with the exchange of primal and dual variables $\bbx_{i,t}$ and $\lam_{ij,t}$ that are needed to implement the primal iteration at time $t$. These local operations repeated in synchrony by all nodes is equivalent to the centralized operations in \eqref{eq:sp_primal}-\eqref{eq:sp_dual}.

We analyze the iterations in \eqref{eq:sp_primal}-\eqref{eq:sp_dual}, which implies convergence of the equivalent iterations in \eqref{eq:local_primal_update} - \eqref{eq:local_dual_update} in the following section. We close here with an example and a remark.

%%%%%%%%%%%%%%%%%%%%%%%%%%%%%%%%%%%%%%%%%%%%%%%%%%%%%%%%%%%%%%%%%%%%%%%%%%%
%%%   E   X   A   M   P   L   E   %%%%%%%%%%%%%%%%%%%%%%%%%%%%%%%%%%%%%%%%%
%%%%%%%%%%%%%%%%%%%%%%%%%%%%%%%%%%%%%%%%%%%%%%%%%%%%%%%%%%%%%%%%%%%%%%%%%%%
%
\medskip\noindent{\bf Example (LMMSE Estimation of a Random Field).} Revisit the random filed estimation problem of Section \ref{sec:prob} that we summarize in the problem formulation in \eqref{eq:dist_rls}. Recalling the identifications $f_i(\bbx_i,\bbtheta_i) = \|\bbH_{i}\bbx_{i} - \bbtheta_{i}\|^2$ and $h_{ij}(\bbx_i,\bbx_j )= (1/2)\|\bbx_{i}- \bbx_{j} \|^2$ it follows that the local primal update in \eqref{eq:local_primal_update} takes the form
\begin{align} \label{eq:rls_primal_i}
   \bbx_{i,t+1} 
      = \ccalP_{\ccalX}\Big[ \bbx_{i,t} - \eps_t  
          \Big[2\bbH_{i}^T\big(\bbH_{i}\bbx_{i,t}-\bbtheta_{i,t}\big)
            + \frac{1}{2}\sum_{j \in n_i} 
              \Big(\lam_{ij,t}+ \lam_{ji,t}\Big)
              \Big(\bbx_{i,t}   - \bbx_{j,t}   \Big) \Big]\Big].
\end{align}
Likewise, the specific form of the dual update in \eqref{eq:local_dual_update} is
\begin{align} \label{eq:quadratic_dual_ij}
   \lam_{ij,t+1}\!\! =\Big[ \! (1 \!-\! \eps_t \delta) \lam_{ij,t} \!+\! (\eps_t/2)\! \!\left(\|\bbx_{i,t} \!- \!\bbx_{j,t} \|^2 \!\!- \!\gamma_{ij} \right)\!\! \Big]_+\; .
\end{align}
The empirical utility of the decentralized estimation scheme in \eqref{eq:rls_primal_i} - \eqref{eq:quadratic_dual_ij} is studied in Section \ref{sec:fields}. Alternative functional forms for the network proximity constraints are studied for a source localization problem in Section \ref{sec:sims} . $\qed$ 

%%%%%%%%%%%%%%%%%%%%%%%%%%%%%%%%%%%%%%%%%%%%%%%%%%%%%%%%%%%%%%%
%%%   R   E   M   A   R   K   %%%%%%%%%%%%%%%%%%%%%%%%%%%%%%%%%
%%%%%%%%%%%%%%%%%%%%%%%%%%%%%%%%%%%%%%%%%%%%%%%%%%%%%%%%%%%%%%%
%
\begin{remark}\normalfont If the proximity constants are $\gamma_{ij} = \gamma_{ji}$ and the initial Lagrange multipliers satisfy $\lambda_{ij,0} = \lambda_{ji,0}$ it follows from \eqref{eq:local_dual_update} that $\lambda_{ij,t} = \lambda_{ji,t}$ for all subsequent times $t$. This is as it should be because the constraints $h_{ij}(\bbx_i, \bbx_{j} ) \leq \gamma_{ij}$ and $h_{ji}(\bbx_j, \bbx_{i} ) \leq \gamma_{ji}$ are redundant. If these multipliers are equal for all times, the primal update in \eqref{eq:local_primal_update} does not necessitate exchange of dual variables. This does not save communication cost as it is still necessary to exchange primal variables $\bbx_{i,t}$.
\end{remark}

%!TEX root = root.tex
\section{Convergence Analysis}\label{sec:convergence}

We turn to establishing that the saddle point algorithm defined by \eqref{eq:sp_primal}-\eqref{eq:sp_dual} converges to the primal-dual optimal point of the problem stated in \eqref{eq:coop_stoch_opt} when a constant algorithm step-size is used. In particular, we establish bounds on the objective function error sequence $F(\bbx_t) - F(\bbx^*)$ and the network-aggregate constraint violation, where $\bbx^*$ is defined by \eqref{eq:coop_stoch_opt}. As a consequence, the time-average primal vector converges to the optimal objective function $F(\bbx^*)$ at a rate of $\ccalO(1/\sqrt{T})$, while incurring constraint violation on the order of $\ccalO(T^{-1/4})$, where $T$ is the total number of iterations. To establish these results, we note some facts of the problem setting, and then introduce a few standard assumptions.

First, observe that the dual stochastic gradient is independent of random variables $\bbtheta_{i,t}$ [cf. \eqref{eq:local_dual_update}], and hence for all $t$, 
\begin{align}\label{eq:dual_avg_grad_equals_dual_stoch_gradient}
{\nabla}_{\bblambda} \ccalL (\bbx_{t},\bblam_t) = {\nabla}_{\bblambda} \hat{\ccalL}_t (\bbx_{t},\bblam_t).
\end{align}
Also pertinent to analyzing the performance of the stochastic saddle point method is the fact that the primal stochastic gradient of the Lagrangian is an unbiased estimator of the true primal gradient. Let $\ccalF_t$ be a sigma algebra that measures the history of the algorithm up until time $t$, i.e. a collection that contains at least the variables $\{\bbx_u, \bblam_u, \bbtheta_u\}_{u=1}^t \subseteq \ccalF_t$. That the primal stochastic gradient is an unbiased estimate of the true primal gradients means that,
\begin{align}\label{eq:unbiased}
   \E{  \nabla_{\bbx} \hat{\ccalL}(\bbx_t, \bblam_t) \given \ccalF_t   } = {\nabla}_\bbx {\ccalL}(\bbx_t, \bblam_t)   \; .  
\end{align}
Furthermore, the compactness of the sets $\ccalX$ permits the bounding of the magnitude of the iterates $\bbx_{i,t}$ by a constant $R/N$, which in turn implies that the network-wide iterates may be bounded in magnitude as
\begin{equation}\label{eq:primal_iterate}
\|\bbx_t \| \leq R \text{ for all } t \; .
\end{equation}
To prove convergence of the stochastic saddle point method, some conditions are required of the network, loss functions, and constraints, which we state below.

%%%%%%%%%%%%%%%%%%%%%%%%%%%%%%%%%%%%%%%%%%%%%%%%%%%%%%%%%%%%%%%%%%%%%%%%%%%
%%%   A   S   S   U   M   P   T   I   O   N   %%%%%%%%%%%%%%%%%%%%%%%%%%%%%
%%%%%%%%%%%%%%%%%%%%%%%%%%%%%%%%%%%%%%%%%%%%%%%%%%%%%%%%%%%%%%%%%%%%%%%%%%%
%
\begin{assumption} \label{as:first} (Network connectivity) The network $\ccalG$ is symmetric and connected with diameter $D$.
\end{assumption}

%%%%%%%%%%%%%%%%%%%%%%%%%%%%%%%%%%%%%%%%%%%%%%%%%%%%%%%%%%%%%%%%%%%%%%%%%%%
%%%   A   S   S   U   M   P   T   I   O   N   %%%%%%%%%%%%%%%%%%%%%%%%%%%%%
%%%%%%%%%%%%%%%%%%%%%%%%%%%%%%%%%%%%%%%%%%%%%%%%%%%%%%%%%%%%%%%%%%%%%%%%%%%
%
\begin{assumption} \label{as:last} (Smoothness) 
The stacked instantaneous objective is Lipschitz continuous in expectation with constant $L_f$, i.e. for distinct primal variables $\bbx, \tbx \in \ccalX$ and all $\bbtheta$, we have
\begin{align} \label{eq:lipschitz}
\E{\| f(\bbx,\bbtheta)\!  -\!  f(\tbx,\bbtheta) \|}  \!&\leq \!L_{ {f}} \|\bbx \!-\! \tbx\| \;, \nonumber \\
\end{align} 
Moreover, the stacked constraint function $h(\bbx)$ is Lipschitz continuous with modulus $L_h$. That is, for distinct primal variables $\bbx, \tbx \in \ccalX$, we may write
\begin{align} \label{eq:lipschitz2}
\| h(\bbx) \! - \! h(\tbx) \| \! &\leq\! L_{ {h}} \|\bbx \!-\! \tbx\| .
\end{align} 
 \end{assumption}
% %%%%%%%%%%%%%%%%%%%%%%%%%%%%%%%%%%%%%%%%%%%%%%%%%%%%%%%%%%%%%%%%%%%%%%%%%%%
%%%%   A   S   S   U   M   P   T   I   O   N   %%%%%%%%%%%%%%%%%%%%%%%%%%%%%
%%%%%%%%%%%%%%%%%%%%%%%%%%%%%%%%%%%%%%%%%%%%%%%%%%%%%%%%%%%%%%%%%%%%%%%%%%%%
%%
%\begin{assumption} \label{as:last} (Bounded Gradients) 
%, the gradients of the stacked instantaneous objective $f$ and constraint function $h$ with respect to $\bbx$ are bounded in expectation by constants $G_{f}$ and $G_{h}$, which implies
%%
%\begin{equation}\label{eq:grad_bound}
%\E{\| {\nabla}_{\bbx}f(\bbx,\bbtheta)\|} \leq G_{\bbx} \; , \  \E{\| {\nabla}_{\bbx}h(\bbx)\|} \leq G_{h} \; .
%\end{equation}
% \end{assumption}
Assumption \ref{as:first} ensures that the graph is connected and the rate at which information diffuses across the network is finite. This condition is standard in distributed algorithms. Assumption \ref{as:last} states that the stacked objective and constraints are sufficiently smooth, and  have bounded gradients. These conditions are common in analysis of descent methods.  

Assumption \ref{as:last} taken with the bound on the primal iterates [cf. \eqref{eq:primal_iterate}] permits the bounding of the expected primal and dual gradients of the Lagrangian by constant terms and terms that depend the magnitude of the dual variable. In particular, we compute the mean-square-magnitude of the primal gradient of the stochastic augmented Lagrangian as
\begin{align}\label{eq:mean_grad_bound_primal}
\mathbb{E}[  \nabla_\bbx \hat{\ccalL} (\bbx, \bblam) \|^2]
%&= \Big\| \sum_{i=1}^N \nabla_{\bbx_i} f_i(\bbx_i, \bbtheta) 
%+ \frac{1}{2} \sum_{i=1}^N \sum_{j \in n_i} \lambda_{ij} \nabla_{\bbx_i} \nabla_{\bbx_i} h_{ij} (\bbx_i, \bbx_j) \Big\|^2 \nonumber \\
&\!\leq \! N \max_i  \E{\| \nabla_{\bbx_i} f_i(\bbx_i, \bbtheta)  \|^2 }  \\
&\quad\!+\! M \|\bblambda\|^2\!\!\max_{(i,j)\in \ccalE }{\|\nabla_{\bbx_i} h_{ij} (\bbx_i, \bbx_j) \|^2} \nonumber \\
&\!\leq\! N L_{ {f}}^2 \!+\!\!M L_{ {h}}^2 {\|\bblam\|^2}\! \!\leq\! (N\!\!+ \!M) L^2(1\!+\! \|\bblam\|^2) \nonumber
\end{align}
where we have applied the triangle inequality, Cauchy-Schwartz in the first expression and considered the worst-case bounds. The second inequality makes use of the smoothness properties defined in \eqref{eq:lipschitz} and the fact that the constraint $h_{ij}(\bbx_i, \bbx_j)$ is independent of $\bbtheta$. On the right-hand side of \eqref{eq:mean_grad_bound_primal} we have defined $L:=\max{}(L_{f}, L_{g})$ to simplify the expression.
 We further may derive a bound on the expected magnitude of the dual stochastic gradient of the augmented Lagrangian by making use of Assumption \ref{as:last}. In particular, we may write
\begin{align}\label{eq:mean_grad_bound_dual}
\!\!\!\!\!\E{ \| \nabla_{\!\bblam} \hat{\ccalL}\! (\bbx, \!\bblam\!) \|^2\! }& \!\! \leq 
\! M\!\! \max_{(i,j)\in\ccalE} \!  (  h_{ij}(\bbx_i, \bbx_j \!)  \!-\!\gamma_{ij}\!)^2 
\! \! +\!  \delta^2\! \eps_t^2 \| \bblambda \|^2   \\
& \!\leq \!\! M \! L_h^2 \|\bbx\|^2 \!\!+ \!\delta^2\!  \eps_t^2 \| \bblambda \|^2 %\nonumber \\
\!\! \leq\! M L_h^2 R^2 \! \!+ \!\delta^2 \! \eps_t^2 \| \bblambda \|^2\! .\nonumber
\end{align}
The first inequality makes use of the triangle inequality and a worst-case bound on the constraint slack, whereas the second uses the Lipschitz continuity of the constraint (Assumption \ref{as:last}), and the last is an application of the compactness of the primal domain $\ccalX^N$. We proceed with a remark and then turn to establishing our main result.
\begin{remark} \normalfont
Rather than bound the primal and dual gradients of the Lagrangian by constants, as is conventionally done in the analysis of primal-dual algorithms, we instead consider upper estimates in terms of the magnitude of the dual variable $\bblam$. In doing so, we alleviate the need for the dual variable to be restricted to a compact subset of the nonnegative real numbers $\reals_{+}^M$. The use of unbounded Lagrange multipliers allow us to mitigate the growth of constraint violation over time using the dual regularization term $(\delta \eps_t/2)\|\bblambda\|^2$ in \eqref{eq:lagrangian}. 
\end{remark}
%
%%%%%%%%%%%%%%%%%%%%%%%%%%%%%%%%%%%%%%%%%%%%%%%%%%%%%%%%%%%%%%%%%%%%%%%%%%%
%%%   M   A   I   N       M   A   T   T   E   R   %%%%%%%%%%%%%%%%%%%%%%%%%
%%%%%%%%%%%%%%%%%%%%%%%%%%%%%%%%%%%%%%%%%%%%%%%%%%%%%%%%%%%%%%%%%%%%%%%%%%%
%
The following lemma is used in the proof of the main theorem, and bounds the Lagrangian difference $\hat{\ccalL}_t (\bbx_{t}, \bblam_{t}) - \hat{\ccalL}_t(\bbx, \bblam_t)$ by a telescopic quantity involving the primal and dual iterates, as well as the magnitude of the primal and dual gradients. 
%Assumption \ref{as:last} establishes that the stochastic gradients of the Lagrangian have finite variance, a standard condition in stochastic optimization which is satisfied in most cases. Assumption \ref{as:optimal_multiplier} simply states that the set $\Lam$ has to be sufficiently large so as to include some of the optimal multipliers.
%
%Before stating our main results, we present a lemma which will subsequently be used in the proof of Theorem \ref{thm1}. 
%%%%%%%%%%%%%%%%%%%%%%%%%%%%%%%%%%%%%%%%%%%%%%%%%%%%%
%%%%%%%%%%%%%%%%%%%%%%%%%%%%%%%%%%%%%%%%%%%%%%%%%%%%%
%%%  L E M M A %%%%%%%%%%%%%%%%%%%%%%%%%%%%%%%%%%%%%%%%%
%%%%%%%%%%%%%%%%%%%%%%%%%%%%%%%%%%%%%%%%%%%%%%%%%%%%%%
\begin{lemma}\label{lemma1}
Denote as $(\bbx_t, \bblam_t )$ the sequence generated by the saddle point algorithm in  \eqref{eq:sp_primal} and \eqref{eq:sp_dual} with stepsize $\eps_t$. If Assumptions \ref{as:first} - \ref{as:last} hold, the instantaneous Lagrangian difference sequence $\hat{\ccalL}_t (\bbx_{t}, \bblam_{t}) - \hat{\ccalL}_t(\bbx, \bblam_t)$ satisfies the decrement property
\begin{align}\label{eq:lemma1}
	\hat{\ccalL}_t & (\bbx_{t}, \bblam) - \!\hat{\ccalL}_t(\bbx, \bblam_t)    \\
	  & \leq \! \frac{1}{2\eps_t} \!\Big(\! \|\bbx_t \!-\! \bbx \|^2 \!-\! \|\bbx_{t+1}\! -\! \bbx \|^2
	   \! +\!  \|\bblam_t - \bblam \|^2\!  -\!  \|\bblam_{t+1}\! - \!\bblam \|^2 \! \Big) \!  +  \frac{ \eps_t  }{2} \Big( \|{\nabla}_{\bbx}\hat{\ccalL}_t(\bbx_t,\bblam_t\!) \|^2 +  \| {\nabla}_{\bblam}\hat{\ccalL}_t(\bbx_{t},\bblam_t) \|^2  \Big)\; . \nonumber
\end{align}
\end{lemma}

\begin{myproof}
See Appendix B.
\end{myproof}

Lemma \ref{lemma1} exploits the fact that the stochastic augmented Lagrangian is convex-concave with respect to its primal and dual variables to obtain an upper bound for the difference $\hat{\ccalL}_t (\bbx_{t}, \bblam_{t}) - \hat{\ccalL}_t(\bbx, \bblam_t)$ in terms of the difference between the primal and dual iterates to a fixed primal-dual pair $(\bbx, \bblam)$ at the next and current time, as well as the square magnitudes of the primal and dual gradients. 
This property is the basis for establishing the convergence of the primal iterates to their constrained optimum given by \eqref{eq:coop_stoch_opt} in terms of objective function evaluation and constraint violation, when a specific constant algorithm step-size is chosen, as we state next.

%. In the following theorem, we use Lemma \ref{lemma1} to establish that after a fixed number $T$ of iterations, the sequence of objective function values produced by the saddle point method with a constant step-size approaches its optimal argument, as well as achieves satisfies tight bounds on the amount of constraint violation up to the present time.
%
%%%%%%%%%%%%%%%%%%%%%%%%%%%%%%%%%%%%%%%%%%%%%%%%%%%%%
%%% T H E O R E M%%%%%%%%%%%%%%%%%%%%%%%%%%%%%%%%%%%%%%%%%
%%%%%%%%%%%%%%%%%%%%%%%%%%%%%%%%%%%%%%%%%%%%%%%%%%%%%%%%%%%%%%%%%%%%%%%%%%%
%
\begin{thm}\label{theorem1}
Denote  $(\bbx_t, \bblam_t )$ as the sequence generated by the saddle point algorithm in  \eqref{eq:sp_primal}-\eqref{eq:sp_dual} and suppose Assumptions \ref{as:first} - \ref{as:last} hold. Suppose the algorithm is run for $T$ iterations with a constant step-size selected as $\eps_t =\eps=1/\sqrt{T}$, then the time aggregation of the objective function error sequence $F(\bbx_t) - F(\bbx^*)$, with $\bbx^*$ defined as in \eqref{eq:coop_stoch_opt}, grows sublinearly with the final iteration index $T$ as 
\begin{align} \label{eq:theorem11}
\sum_{t=1}^T [F(\bbx_t) \!- \! F(\bbx^*)] 
\!& \leq    \ccalO(\sqrt{T}).
\end{align}
Moreover, the time-aggregation of the constraint violation of the algorithm grows sublinearly final time $T$ as 
\begin{align} \label{eq:theorem12}
\sum_{(i,j)\in\ccalE}\!\Big[\!\sum_{t=1}^T \Big(
 h_{ij}(\bbx_{i,t},\bbx_{j,t} ) -  \gamma_{ij} 
\!\!\Big) \Big]_+
\leq \ccalO(T^{3/4}).
\end{align}
\end{thm}
%%%%%%%%%%%%%%%%%%%%%%%%%%%%%%%%%%%%%%%%%%%%%%%%%%%%%%%%%%%%%%%%%%%%%%%%%%%
%%%   P   R   O   O   F   %%%%%%%%%%%%%%%%%%%%%%%%%%%%%%%%%%%%%%%%%%%%%%%%%
%%%%%%%%%%%%%%%%%%%%%%%%%%%%%%%%%%%%%%%%%%%%%%%%%%%%%%%%%%%%%%%%%%%%%%%%%%%
%
\begin{myproof} 
We first consider the expression in \eqref{eq:lemma1}, and expand the left-hand side using the definition of the augmented Lagrangian in \eqref{eq:stoch_lagrangian}. Doing so yields the following expression,
\begin{align} \label{eq:stoch_lagrangian_expand}
& \sum_{i=1}^N  [f_i(\bbx_{i,t},\bbtheta_{i,t}) 
 - f_i(\bbx_{i},\bbtheta_{i,t})]  
  +\frac{\delta \eps_t}{2}( \| \bblambda_t\|^2\! -\! \|\bblambda\|^2) \\
&\quad\!\!  +\!\!\!\!\sum_{(i,j)\in\ccalE}\!\!\!\! \left[
\lambda_{ij} \left(h_{ij}(\bbx_{i,t},\bbx_{j,t} )\! -\! \gamma_{ij} \right)
\!-\! \lambda_{ij,t} \left(h_{ij}(\bbx_{i},\bbx_{j} )\! -\! \gamma_{ij} \right) \right] \nonumber\\
&\quad\leq   
 \frac{1}{2\eps_t} \Big(\! \|\bbx_t \!-\! \bbx \|^2\! \!-\! \|\bbx_{t+1} \!-\! \bbx \|^2\!
	    +  \|\bblam_t \!-\! \bblam \|^2 \!\! -  \!\|\bblam_{t+1} \!-\! \bblam \|^2 \! \Big)  \nonumber\\
&	\qquad +  \frac{ \eps_t  }{2} \Big( \|{\nabla}_{\bbx}\hat{\ccalL}_t(\bbx_t,\bblam_t\!) \|^2 +  \| {\nabla}_{\bblam}\hat{\ccalL}_t(\bbx_{t},\bblam_t) \|^2  \Big) ,\nonumber
\end{align}
after gathering like terms. Compute the expectation of \eqref{eq:stoch_lagrangian_expand} conditional on $\ccalF_0$ and substitute in the bounds for the mean-square-magnitude of the primal and dual gradients of the stochastic augmented Lagrangian given in \eqref{eq:mean_grad_bound_primal} and \eqref{eq:mean_grad_bound_dual}, respectively, into the right-hand side to obtain
\begin{align} \label{eq:lagrangian_expand}
F&(\bbx_t) -  F(\bbx)
  +\frac{\delta \eps_t}{2}( \| \bblambda_t\|^2\! -\! \|\bblambda\|^2) \\
& +\!\!\!\!\sum_{(i,j)\in\ccalE}\!\!\!\! \left[
\lambda_{ij} \left(h_{ij}(\bbx_{i,t},\bbx_{j,t} )\! -\! \gamma_{ij} \right)
\!-\! \lambda_{ij,t} \left(h_{ij}(\bbx_{i},\bbx_{j} )\! -\! \gamma_{ij} \right) \right] \nonumber\\
&\leq   
\! \frac{1}{2\eps_t}\! \Big(\! \|\bbx_t \!-\! \bbx \|^2\! -\! \|\bbx_{t+1} \!-\! \bbx \|^2
	    \!\!+\! \! \|\bblam_t \!-\!\! \bblam \|^2 \!\! -  \!\|\bblam_{t+1} \!-\! \bblam \|^2 \! \Big)  \nonumber\\
& \quad	 +  \frac{ \eps_t  }{2} \Big((N+ M) L^2(1+ \|\bblam_t\|^2) + M L_h^2 R^2 \! \!+ \!\delta^2 \! \eps_t^2 \| \bblambda_t \|^2 \Big)\; , \nonumber
\end{align}
where we have also used the fact that the constraint functions $h_{ij}(\bbx_i, \bbx_j)$ appearing as the third term on the left-hand side are independent of $\bbtheta$, and noting that the right-hand side of \eqref{eq:lagrangian_expand} is equal to its expectation. Observe that $\bbx\in \ccalX$ is an arbitrary feasible point, which implies that $h_{ij}(\bbx_{i},\bbx_{j} ) \leq \gamma_{ij}$ for all $(i,j)\in \ccalE$. Making use of this property to annihilate the last term on the left-hand side of \eqref{eq:lagrangian_expand} and subtracting $(\delta \eps_t/2 )\|\bblam_t\|^2$ from both sides yields
\begin{align} \label{eq:lagrangian_feasibility}
&F(\bbx_t) -  F(\bbx)
 +\!\!\!\!\sum_{(i,j)\in\ccalE}\!\!\left[
\lambda_{ij} \left(h_{ij}(\bbx_{i,t},\bbx_{j,t} )\! -\! \gamma_{ij} \right)
 - \frac{\delta \eps_t}{2} \lambda_{ij}^2 \right] \nonumber\\
&\quad \leq    
\! \frac{1}{2\eps_t}\! \Big(\! \|\bbx_t \!-\! \bbx \|^2\! -\! \|\bbx_{t+1} \!-\! \bbx \|^2
	    +  \|\bblam_t \!-\! \bblam \|^2 \! -  \!\|\bblam_{t+1} \!-\! \bblam \|^2 \! \Big)  \nonumber\\
&\qquad	 \!+ \! \frac{ \eps_t  }{2}\! \Big( K %\nonumber\\
+\! ((N\!+\!\!M) L^2\!\!\! +\! \delta^2 \eps_t^2 \!- \!\delta )\|\bblam_t\|^2 \! \Big). 
\end{align}
after reordering terms, and defining the constant $K:=(N +M) L^2 + M L_h^2 R^2$. Now sum the expression \eqref{eq:lagrangian_feasibility} over times $t=1,\dots, T$ for a fixed $T$, and select the constant $\delta$  to satisfy $ (N+M) L^2 + \delta^2 \eps_t^2 \leq \delta $ for a constant step-size $\eps_t=\eps$ to drop the term involving $\|\bblam_t\|^2$ from the right-hand side as
\begin{align} \label{eq:sum_T}
\sum_{t=1}^T [F(\bbx_t \!) \!- \!\! F(\bbx\!)] 
\! &+\!\!\!\!\!\!\sum_{(i,j)\in\ccalE}\!\!\!\!\lambda_{ij}\! \Big[\!\!\sum_{t=1}^T
\left(h_{ij}(\bbx_{i,t},\bbx_{j,t} \!)\! -\! \gamma_{ij}\! \right)
\!\!\Big]\! \!- \! \!\frac{\delta \eps T}{2} \|\bblambda\|^2  \nonumber\\
& \leq    
\! \frac{1}{2\eps}\! \Big(\! \|\bbx_1 \!-\! \bbx \|^2\!
	   \!+\!  \|\bblam_1 \!-\! \bblam \|^2 \! \Big)  
	 +  \frac{ \eps K  }{2}.
\end{align}
In \eqref{eq:sum_T}, we exploit the telescopic property of the summand over differences in the magnitude of primal and dual iterates to a fixed primal-dual pair $(\bbx, \bblam)$ which appears as the first term on right-hand side of \eqref{eq:lagrangian_feasibility}. By assuming the dual variable is initialized as $\bblam_1=\bb0$ and subtracting the resulting $(1/2\eps)\|\bblam\|^2$ term to the other side, the expression in \eqref{eq:sum_T} becomes
\begin{align} \label{eq:sum_T_dual_initialize}
\sum_{t=1}^T& [F(\bbx_t) \!- \! F(\bbx)] 
\! +\!\!\!\sum_{(i,j)\in\ccalE}\!\lambda_{ij} \Big[\!\sum_{t=1}^T \left(
 h_{ij}(\bbx_{i,t},\bbx_{j,t} )\! -\! \gamma_{ij} \right) \Big] \nonumber \\
 &- \Big(\!\frac{\delta \eps T}{2} \!+\! \frac{1}{2\eps}\Big)\! \|\bblambda\|^2 
 \leq    \frac{1}{2\eps} \|\bbx_1 \!-\! \!\bbx \|^2
	 +  \frac{ \eps T K}{2}. 
\end{align}
At this point, we note that the left-hand side of the expression in \eqref{eq:sum_T_dual_initialize} consists of two terms. The first is the accumulation over time of the global loss, which is a sum of all local losses at each node as defined in \eqref{eq:dist_stoch_opt}. The second term is the inner product of the an arbitrary Lagrange multiplier $\bblam$ with the time-aggregation of constraint violation, and the last is a term which depends on the magnitude of this multiplier. We may use these later terms to define an ``optimal" Lagrange multiplier to control the growth of the long-term constraint violation of the algorithm. To do so, define the \emph{augmented} dual function $\tilde{g}(\bblam)$ using the later two terms on the left-hand side of \eqref{eq:sum_T_dual_initialize}
\begin{align} \label{eq:augmented_dual_func}
\tilde{g}(\bblam)\!=\!\!\!\!\!\!\sum_{(i,j)\in\ccalE}\!\!\!\lambda_{ij} \Big[\!\sum_{t=1}^T \Big(
\!\! \left(h_{ij}(\bbx_{i,t},\bbx_{j,t} )\! -\! \gamma_{ij} \right)
\!\!\Big) \!\Big]
\!\!-\!\! \Big(\!\frac{\delta \eps T}{2} \!+\! \frac{1}{2\eps}\!\Big)\! \|\bblambda\|^2.
\end{align}
Computing the gradient of \eqref{eq:augmented_dual_func} and solving the resulting stationary equation over the range $\reals_{+}^M$ yields
\begin{align} \label{eq:lambda_tilde}
\tilde{\lambda}_{ij}= \Big(\frac{1}{2 ( T \delta \eps + 1/\eps)}\Big) \!\sum_{t=1}^T \Big(
 h_{ij}(\bbx_{i,t},\bbx_{j,t} )\! -\! \gamma_{ij} 
\!\!\Big)_+
\end{align}
for all $(i,j)\in\ccalE$. Substituting the selection $\bblam=\tilde{\bblam}$ defined by \eqref{eq:lambda_tilde} into \eqref{eq:sum_T_dual_initialize} results in the following expression
\begin{align} \label{eq:lambda_tilde_substitute}
\sum_{t=1}^T [F(\bbx_t) \!- \! F(\bbx)] 
\! +\sum_{(i,j)\in\ccalE}&\!\frac{\Big[\!\sum_{t=1}^T \Big(
 h_{ij}(\bbx_{i,t},\bbx_{j,t} )\! -\! \gamma_{ij} 
\!\!\Big) \Big]^2_+}{2 ( T \delta \eps + 1/\eps) }
\nonumber \\
&\quad \leq    
\! \frac{1}{2\eps} \|\bbx_1 \!-\! \!\bbx \|^2
	 +  \frac{ \eps T K }{2}\! .
\end{align}
Now select the constant step-size $\eps = 1/\sqrt{T}$, and substitute the result into \eqref{eq:lambda_tilde_substitute}, using the formula for $K$ defined following expression \eqref{eq:lagrangian_feasibility}, to obtain
\begin{align} \label{eq:lambda_tilde_substitute_stepsize}
\sum_{t=1}^T &[F(\bbx_t) \!- \! F(\bbx)] 
\! +\sum_{(i,j)\in\ccalE}\!\frac{\Big[\!\sum_{t=1}^T \Big(
 h_{ij}(\bbx_{i,t},\bbx_{j,t} )\! -\! \gamma_{ij} 
\!\!\Big) \Big]^2_+}{2 \sqrt{T} (  \delta + 1) }
\nonumber \\
& \leq    
\! \frac{\sqrt{T}}{2}\!\!\left( \|\bbx_1 \!-\! \!\bbx \|^2 \! \!
	 +   \!(\!N\!+ \!M) L^2 \!\!+ \! M L_h^2 R^2 \right).% 
\end{align}
The expression in \eqref{eq:lambda_tilde_substitute_stepsize} allows us to derive both the convergence of the global objective and the feasibility of the stochastic saddle point iterates. 

We first consider the objective error sequence $F(\bbx_t) - F(\bbx^*)$. To do so, subtract the last term on the left-hand side of \eqref{eq:lambda_tilde_substitute_stepsize} from both sides, and note that the resulting term is non-positive. This observation allows us to omit the constraint slack term in \eqref{eq:lambda_tilde_substitute_stepsize}, which taken with the selection $\bbx=\bbx^*$ [cf. \eqref{eq:coop_stoch_opt}], yields
\begin{align} \label{eq:objective_convergence}
\sum_{t=1}^T [F(\bbx_t) \!- \! F(\bbx^*)] 
\!\! \leq    
\! \frac{\sqrt{T}}{2}\!\!\left( \|\bbx_1 \!-\! \!\bbx^* \|^2 \! \!
	 +   K \right) = \ccalO(\sqrt{T}) \; ,
\end{align}
%
%The objective error sequence $F(\bbx_t ) - F(\bbx^*)$ in \eqref{eq:objective_convergence} allows us to establish that $\liminf_t a_t $ = 0. 
%
%
which is as stated in \eqref{eq:theorem11}. Now we turn to establishing a sublinear growth of the constraint violation in $T$, using the expression in \eqref{eq:lambda_tilde_substitute_stepsize}. First, observe that the objective function error sequence is bounded above as
\begin{align} \label{eq:objective_upper_bound}
F(\bbx_t) - F(\bbx^*) \leq L_f \|\bbx_t - \bbx^*\| \leq 2 L_f R
\end{align}
Substituting \eqref{eq:objective_upper_bound} into the first term on the left-hand side of \eqref{eq:lambda_tilde_substitute_stepsize} and subtracting the result from both sides yields
\begin{align} \label{eq:feasibility1}
\!\!\sum_{(i,j)\in\ccalE}\!\!\!\!\!\!\frac{\Big[\!\sum_{t=1}^T \!\!\Big(\!
 h_{ij}(\!\bbx_{i,t},\!\bbx_{j,t} \!)\! -  \!\gamma_{ij} 
\!\!\Big)\! \Big]^2_+}{2 \sqrt{T} (  \delta + 1) }
% \\
%&\qquad\qquad\qquad\qquad
\!\! \leq    \!
\! \frac{\sqrt{T}}{2}\!\!\left( \|\bbx_1 \!-\!\! \bbx \|^2 \! \!\!
	 + \! \!K \right) \!\!-\! 2 T\! L_f \! R . 
\end{align}
which, after multiplying both sides by $2 \sqrt{T} (  \delta + 1)$ and computing square-roots, yields
\begin{align} \label{eq:feasibility2}
&\sum_{(i,j)\in\ccalE}\!\Big[\!\sum_{t=1}^T \Big(
 h_{ij}(\bbx_{i,t},\bbx_{j,t} ) -  \gamma_{ij} 
\!\!\Big) \Big]_+
 \\
&\!\!\! \leq    \! \!
\Big[\!\Big(2 \sqrt{T} (  \delta + 1) \! \Big)\! \Big(\!\frac{\sqrt{T}}{2}\!\!\left( \|\bbx_1 \!-\! \!\bbx \|^2 \! \!
	 +  \! \!K \right) %\nonumber \\
	 \!-\!2 T L_f R\!\Big)\! \Big]^{1/2} 
	\!\!\!= \!\ccalO(T^{3/4}). \nonumber
\end{align}
as claimed in \eqref{eq:theorem12}.
\end{myproof} 

Theorem \ref{theorem1} establishes that the stochastic saddle point method, when run with a fixed algorithm step-size, yields an objective function error sequence whose difference is bounded by a constant strictly less times than $T$, the final iteration index. Moreover, the time-accumulation of the constraint violation incurred by the algorithm is strictly smaller than $T$, the final iteration index. Thus, for larger $T$, the time-average difference between $F(\bbx_t)$ and $F(\bbx^*)$ goes to null, as does the average constraint violation.
Theorem \ref{theorem1} also allows us to establish convergence of the average iterates to a specific level of accuracy dependent on the total number of iterations $T$, as we subsequently state.
%%%%%%%%%%%%%%%%%%%%%%%%%%%%%%%%%%%%%%%%%%%%%%%%%%%%%
%%%C O R O L L A R Y%%%%%%%%%%%%%%%%%%%%%%%%%%%%%%%%%%%%%%%%%
%%%%%%%%%%%%%%%%%%%%%%%%%%%%%%%%%%%%%%%%%%%%%%%%%%%%%%%%%%%%%%%%%%%%%%%%%%%
%
\begin{corollary}\label{corollary2}
Let $\bar{\bbx}_T = (1/T)\sum_{t=1}^T \bbx_t$  be the vector formed by averaging the primal iterates $\bbx_t$ over times $t=1,\dots,T$.. Under Assumptions \ref{as:first} - \ref{as:last}, with constant algorithm step-size $\eps_t =1/\sqrt{T}$, the objective function evaluated at $\bar{\bbx}_T$ satisfies 
\begin{align} \label{eq:corollary21}
F(\bar{\bbx}_T) -F(\bbx^*) \leq \ccalO(1/\sqrt{T})
\end{align}
Moreover, the constraint violation evaluated at the average vector $\bar{\bbx}_T$ satisfies
\begin{align} \label{eq:corollary22}
 \sum_{(i,j)\in\ccalE} \Big[
 h_{ij}(\bar{\bbx}_{i,T},\bar{\bbx}_{j,T} ) -  \gamma_{ij} \Big]_+ =\ccalO(T^{-\frac{1}{4}}).
\end{align}
\end{corollary}
%%%%%%%%%%%%%%%%%%%%%%%%%%%%%%%%%%%%%%%%%%%%%%%%%%%%%%%%%%%%%%%%%%%%%%%%%%%
%%%   P   R   O   O   F   %%%%%%%%%%%%%%%%%%%%%%%%%%%%%%%%%%%%%%%%%%%%%%%%%
%%%%%%%%%%%%%%%%%%%%%%%%%%%%%%%%%%%%%%%%%%%%%%%%%%%%%%%%%%%%%%%%%%%%%%%%%%%
\begin{myproof}
Consider the expressions in Theorem \ref{theorem1}. In particular, to prove \eqref{eq:corollary21}, we consider the expression in \eqref{eq:theorem11}, divide the expression by $T$, and use the definition of convexity of the expected objective $F$ and constraint functions $h_{ij}(\bbx_i, \bbx_i)$, which says that the average of function values upper bounds the function evaluated at the average vector, i.e.
\begin{equation}\label{eq:cvx_average}
F(\bar{\bbx}_T)\! \leq \!\frac{1}{T}\!\!\sum_{t=1}^T F(\bbx_t) \; ,  h_{ij}(\bar{\bbx}_{i,T}, \bar{\bbx}_{j,T}) \!\leq\! \frac{1}{T}\!\!\sum_{t=1}^T h_{ij}(\bar{\bbx}_{i,t}, \bar{\bbx}_{j,t})
\end{equation}
Applying the relation \eqref{eq:cvx_average} to the expressions in \eqref{eq:theorem11} and \eqref{eq:theorem12} divided by $T$ yields \eqref{eq:corollary21} and \eqref{eq:corollary22}, respectively.
\end{myproof}

Corollary \ref{corollary2} shows that the average saddle point primal iterates $\bar{\bbx_t}$ converge to within a margin $\ccalO(1/\sqrt{T})$ in terms of objective function evaluation to the optimal objective $F(\bbx^*)$, where $T$ is the number of iterations. Moreover, the primal average vector also achieves yields the bound on the network proximity constraint violation as $\ccalO(T^{-1/4})$. To summarize, when a constant step-size is used, with increasing $T$ we have the following relations: the instantaneous iterates converge on average, whereas the average iterates converge to the optimal objective. This pattern also holds in terms of the algorithm's feasibility performance -- the time average constraint violation approaches null with increasing $T$, whereas the constraint violation of the average vector approaches null. In the next sections, we numerically analyze the performance of the saddle point method for a couple decentralized sequential estimation problems, illustrating their practical utility.
%%%%%%%%%%%%%%%%%%%%%%%%%%%%%%%%%%%%%%%%%%%%%%%%%%%%%%%%%%%%%%%%%%
%%%   F I G U R E   %%%%%%%%%%%%%%%%%%%%%%%%
%%%%%%%%%%%%%%%%%%%%%%%%%%%%%%%%%%%%%%%%%%%%%%%%%%%%%%%%%%%%%%%%%%
%
\begin{figure*}
\setcounter{subfigure}{0}
\begin{subfigure}{.5\columnwidth}
\includegraphics[width=\linewidth, height = 0.55\linewidth]
                {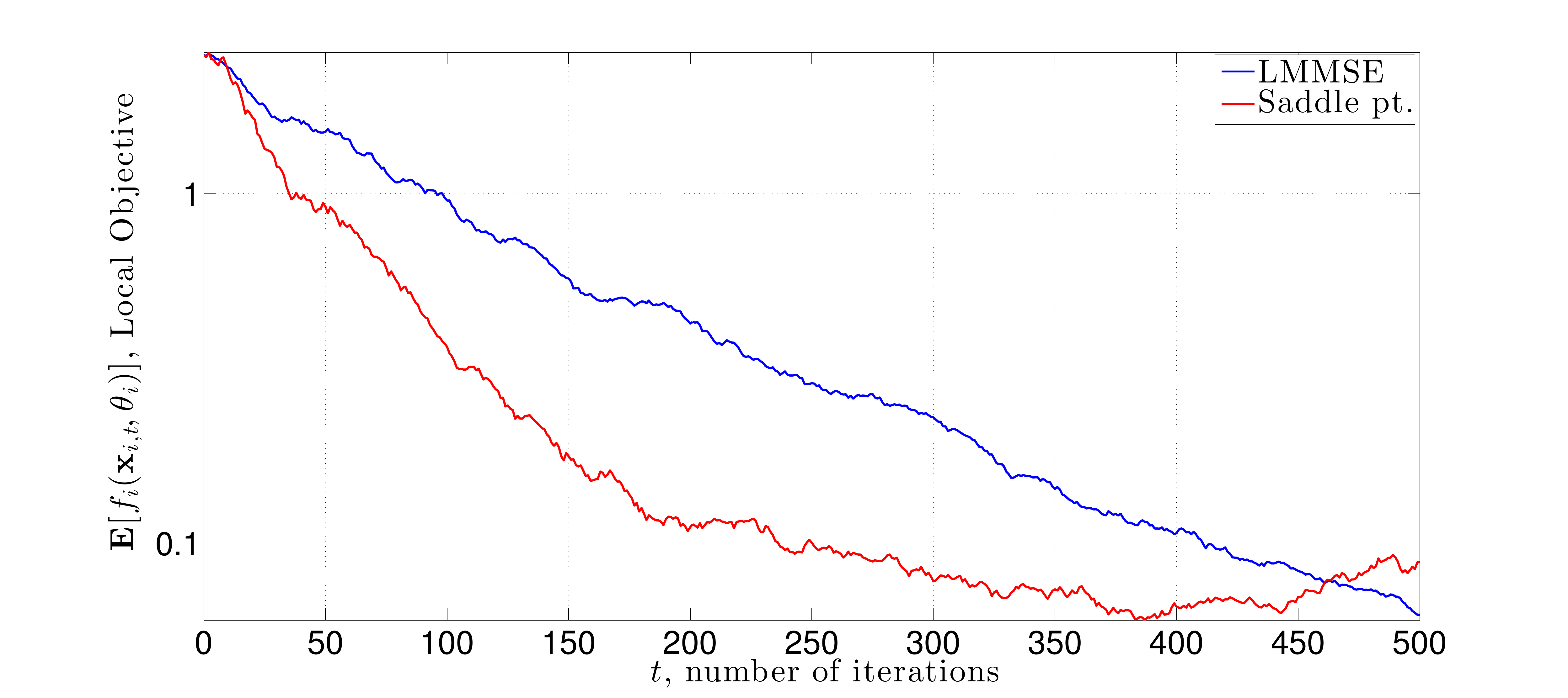}
\caption{Local objective vs. iteration $t$}
\label{subfig:objective}
\end{subfigure}
\begin{subfigure}{.5\columnwidth}
\includegraphics[width=\linewidth,height = 0.55\linewidth]
                {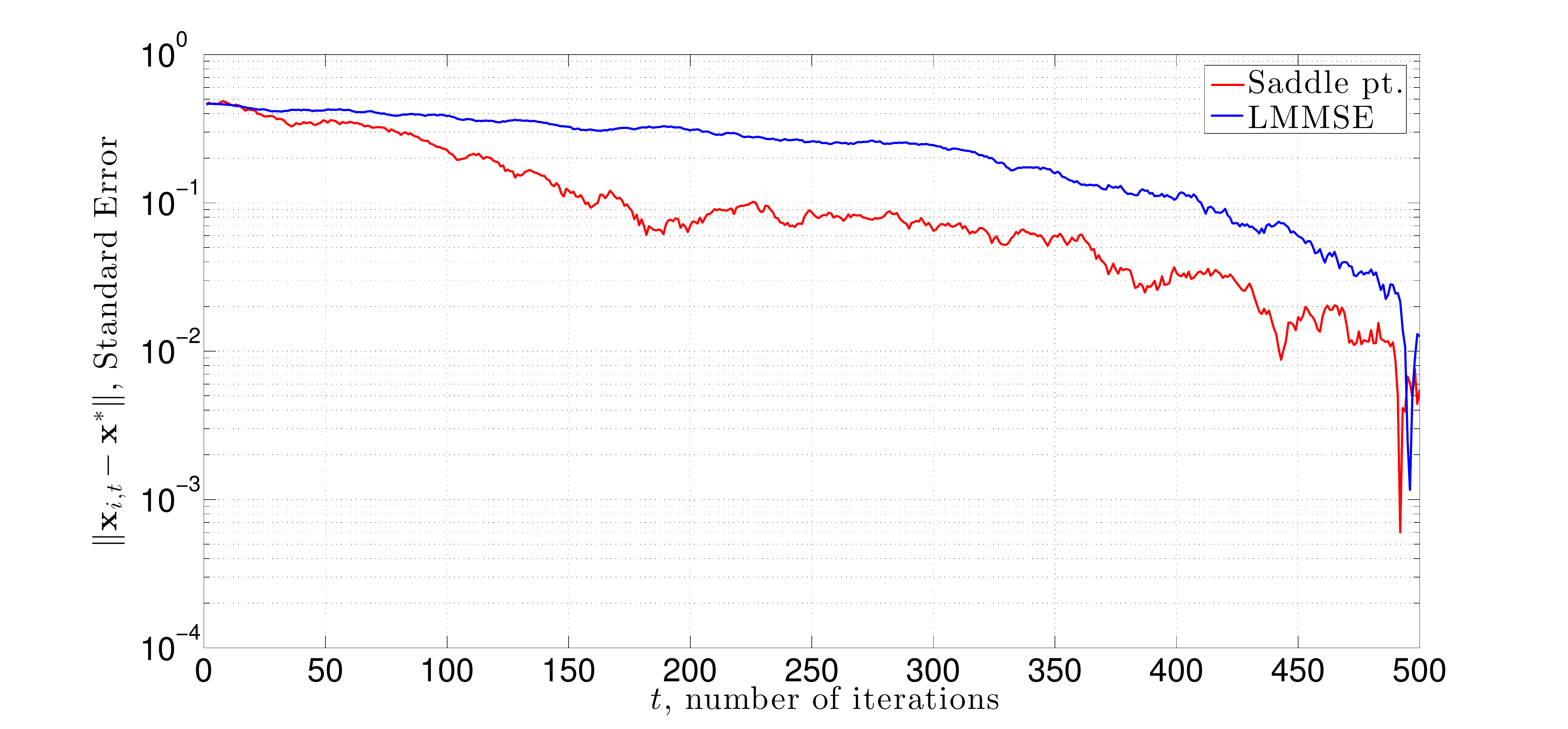}
\caption{Standard error vs. iteration $t$}
\label{subfig:standard_err}
\end{subfigure}
%
%\begin{subfigure}{.693\columnwidth}
%\includegraphics[width=\linewidth,height = 0.6\linewidth]
%                {constraint.pdf}
%\caption{Network Proximity vs. iteration $t$}
%\label{subfig:proximity}
%\end{subfigure}
%%
\caption{Saddle point algorithm applied to the problem of estimating a correlated random field. Nodes are deployed uniformly in a square region of size $200 \times 200$ squared meters in a grid formation, and node estimators are correlated according to the distance-based model $\rho(\bbx_i, \bbx_j)= e^{-\|l_i - l_j\|}$, where $\bbx_i$ and $\bbx_j$ are the decisions of nodes $i$ and $j$, and $l_j$ are their respective locations. Individual sensors learn global information while remaining close to nodes whose information they deem important. Exploiting the correlation structure of the field yields a reduction in the local estimation error and distance to the optimal LMMSE estimate. }\label{fig:fields}\vspace{-3mm}
\end{figure*}

\section{Random Field Estimation}\label{sec:fields}

Consider the task of estimating a spatially correlated random field in a specified region $\ccalA\subset \reals^p$ by making use of a sensor network that we discussed in Sections \ref{sec:prob} and \ref{sec:algorithm}. Interconnected sensors collect observations $\bbtheta_{i,t}$ which are noisy linear transformations of the signal $\bbx\in \reals^p$ they would like to estimate, which are related through the observation model $\bbtheta_{it}=\bbH_{i}\bbx+ \bbw_{i,t}$ with Gaussian noise $\bbw_{i,t} \sim \ccalN(0, \sigma^2 \bbI_q)$   i.i.d across time and node with $\sigma^2 = 2$. The random field is parameterized by the correlation matrix $\bbR_x$, which is assumed to follow a spatial correlation structure of the form $\rho(\bbx_i, \bbx_j)= e^{-\|l_i - l_j\|}$, where $l_i\in \ccalA$ and $l_j\in \ccalA$ are the respective locations of sensor $i$ and sensor $j$ in the deployed region, see, e.g., \cite{1365116}. Observe that now each node has a unique signal-to-noise ratio based upon its location and that information received at more distant nodes are less important; however, their contribution to the aggregate objective $F(\bbx)$ still incentivizes global coordination. This task may be thought of as an online maximum likelihood estimation problem where the estimators of distinct sensors depend on one another.

To solve this problem, we deploy $N=50$ sensors in a grid formulation, where neighboring nodes have a constant distance apart from one another in a  $200\times 200$ meter square region. We make use of the saddle point algorithm [cf. \eqref{eq:local_primal_update} - \eqref{eq:local_dual_update}], whose updates for the random field estimation problem is given by the explicit expressions in \eqref{eq:rls_primal_i} and \eqref{eq:quadratic_dual_ij}, respectively. We select $\gamma_{ij} = \rho(\bbx_i, \bbx_j)$.
Besides the local and global losses which on average converge to a neighborhood of the constrained optima depending on the final iteration index $T$ when a constant step-size is used (Theorem \ref{theorem1}), we also study the standard error to the LMSE estimator $\bbx^*$, i.e. $\| \bbx_{i,t} - \bbx^* \|$. 

To compute $\bbx^*$ for a single time slot, stack observations $\bbtheta=[ \bbtheta_{1} ; \cdots ; \bbtheta_{N} ]$ and observation models $\bbH=[ \bbH_{1} ; \cdots ; \bbH_{N} ]$. Then the least mean squared error (LMSE) estimator for a single time slot of this problem is 
$\bbx^*= (\bbH \bbR_x \bbH^T + \frac{1}{\sigma^2} \bbI )^{-1} \frac{1}{\sigma^2} \bbI \bbtheta$.
To compute the benchmark LMSE $\bbx^*$ for a given run, we stack signals $\bbtheta_{i,t}$ for all nodes $i$ and times $t$ at a centralized location into one large linear system and substitute the sample variance $\hat{\sigma}^2$ in the prior computation.
%
%As a consequence of Thoerems \ref{thm1} and \ref{thm2}, asymptotically we expect the constraint violation to also converge to null, although it is omitted here for clarify. 

%\blue{What are the noise levels?} 
We consider problem instances where observations and signal estimates are scalar ($p=q=1$), the scalar $\bbH=1$, and the a priori signal $\bbx = \bbone$ is set a vector of ones, and run the algorithm for for $T=500$ iterations with a hybrid step-size strategy which is constant for the first $t_0$ iterations and then attenuates, i.e. $\eps_t = \min(\eps, \eps t_0 / t)$ with $t_0=100$  and $\eps=10^{-2}$. We further select the dual regularization parameter $\delta=10^{-5}$. The noise level is set to $\sigma^2=10$. 
%
%The optimal $\bbx^*$ is computed by solving the normal equations of the form $ ( \bbR_x  +\sigma^2 \bbI )\bbx^*= \frac{1}{\sigma^2} \bbI \bbtheta_t$ stacked over all $t$. 
%
We compare the performance of the algorithm with that of a simple LMMSE estimator strategy which does not take advantage of the correlation structure of the sensor network.

In Figure \ref{fig:fields}, we plot the results of this numerical experiment. Figure \ref{subfig:objective} shows the local objective $ {\mathbb E}_{\bbtheta_i}[f_i(\bbx_{i,t},\bbtheta_i)]$ of an arbitrarily chosen node $i\in V$ versus iteration $t$. We observe the numerical behavior of the global objective is similar to the local objective, and is thus omitted. We see that when nodes incorporate the correlation structure of the random field into their estimation strategy via the quadratic proximity constraint with $\gamma_{ij}$ chosen according to the correlation of node $i$ and its neighbors $j\in n_i$, the estimation performance improves. In particular, to achieve the benchmark $ {\mathbb E}_{\bbtheta_i}[f_i(\bbx_i,\bbtheta_i)] \leq .1$, we require $t=247$ versus $t=411$ iterations respectively for the case that the correlation structure is exploited via the saddle point algorithm as compared with a simple LMMSE scheme. We observe that for small $t$ the gain is substantial, but for large $t$ the performance is comparable to the LMMSE strategy.

This improved estimation performance is corroborated in the plot of the standard error $\| \bbx_{i,t} - \bbx^* \|$ to the optimal estimator as compared with iteration $t$ in Figure \ref{subfig:standard_err}. We see that to achieve the benchmark $\| \bbx_{i,t} - \bbx^* \| \leq .1$, the saddle point algorithm requires $t=157$ iterations as compared with $t=414$ for the LMMSE estimator, more than twice as many. We have observed that the benefit of using the saddle point method as compared with simple LMMSE is more substantial in problem instances where the signal to noise ratio is low, and the region $\ccalA$ is larger. In the next section, we study the use of the saddle point method for solving the problem of deploying sensors in a spatial region to locate the position of a source signal.

%In Figure \ref{subfig:proximity}, we plot the local constraint violation $\NP(\bbx_{i,t})$ of an arbitrarily chosen sensor $i\in V$, and observe that the algorithm successsfully keeps the estimates of node $i$ close to those of its neighbors, where the closeness constraint is given by the correlation structure of the random field In particular, for all $t$, we have $\NP(\bbx_{i,t}) \leq 2$, ensuring that the network proximity constraint is satisfied. In doing so, individual sensors are successfully able to incorporate spatial information about the random field into their estimation.

%!TEX root = root.tex
\section{Source Localization}\label{sec:sims}

We now consider the use of the stochastic saddle point method given in \eqref{eq:sp_primal} - \eqref{eq:sp_dual} to solve an online source localization problem. In particular, we consider an array of $N$ sensors, where $\bbl_i \in \reals^p$ denotes the position of the sensor $i$ in some deployed environment $\ccalA\subset \reals^p$. Each node seeks to learn the location of a source signal $\bbx \in \reals^p$ through its access to noisy range observations of the form 
%
%\begin{equation}\label{eq:range}
$r_{i,t} = \| \bbx - \bbl_i \| + \varepsilon_{i,t}$
%\end{equation}
%
where $\varepsilon_t=[\varepsilon_{1,t}; \cdots; \varepsilon_{N,t}]$ is some unknown noise vector. The goal of each sensor $i$ in the network is, given access to sequentially observed range measurements $r_{i,t}$, to learn the position of the source $\bbx$, assuming it is aware of its location $\bbl_i$ in the deployed region. Range-based source localization has been studied in a variety of fields, from wireless communications to geophysics \cite{5413253,2011AGUFM.S31C2249R}.

Rather than considering a range-based least squares problem, which is nonconvex and may be solved approximately using semidefinite relaxations \cite{1326215}, we consider the squared range-based least squares (SR-LS) problem, stated as
\begin{alignat}{2} \label{eq:sr_ls}
\bbx^*:=\ 
&\argmin_{\bbx\in \reals^p }\ &&\sum_{i=1}^N {\mathbb E}_{\bbr_i}\Big(\|\bbl_{i} - \bbx\|^2 - r_i^2 \Big)^2 \; . %\\
% &\text{\ s.t. } \ &&(1/2)\|\bbx_{i}- \bbx_{j} \|^2 \leq \gamma_{ij}, \quad\text{ for all }j\in n_i. \nonumber
\end{alignat}
 Although this problem is also nonconvex, it may be solved approximately in a lower-complexity manner as a quadratic program -- see \cite{4472183}, Section II-B and references therein. To do so, expand the square in the first term in the objective stated \eqref{eq:sr_ls} and consider the modified argument inside the expectation $(\alpha - 2 \bbl_{i}^T \bbx + \|\bbl_i \|^2 - r_i^2 )^2$
with the constraint $\|\bbx \| =\alpha$. Proceeding as in \cite{4472183}, Section II-B, approximate this transformation by a convex unconstrained problem by defining matrix $\bbA \in \reals^{N \times (p+1)}$ whose $i$th row associated with sensor $i$ is given as $\bbA_i = [-2 \bbl_i^T ; 1]$, and vector $\bbb\in \reals^N$ with $i$th entry as $\bbb_i = r_i^2 - \|\bbl_i \|^2 $ and relaxing the constraint $\|\bbx \| =\alpha$. Further define $\bby = [\bbx ; \alpha]\in \reals^{p+1}$, which allows \eqref{eq:sr_ls} to be approximated as
\begin{alignat}{2} \label{eq:sr_ls_relax}
\bby^*:=\ 
&\argmin_{\bby\in \reals^{p+1} }\ && \sum_{i=1}^N \mathbb{E}_{\bbb_i} \Big( \|\bbA_i \bby - \bbb_i \|^2\Big) \; ; %\\
\end{alignat}
which is a least mean-square error problem. We note that the techniques in \cite{4472183} to solve this problem exactly do not apply to the online setting\cite{Moré93generalizationsof}.
 
We propose solving \eqref{eq:sr_ls_relax} in decentralized settings which more effectively allow for each sensor to operate based on real-time observations. To do so, each sensor keeps a local copy $\bby_i$ of the global source estimate $\bby$ based on information that is available with local information only and via message exchange with neighboring sensors. However, each sensor would still like to attain the greater estimation accuracy associated with aggregating range observations over the entire network. We proceed to illustrate how this may be achieved by using the proximity constrained optimization in Section \ref{sec:prob}.

In application domains such as wireless communications or acoustics \cite{1458275}, the quality of the observed range measurements is better for sensors which are in closer proximity to the source. Motivated by this fact, we consider the case where sensor $i$ weights the importance of neighboring sensors $j\in n_i$ by aiming to keep its estimate $\bbx_i$ within an $\ell_2$ ball centered at its neighbors estimate $\bbx_j$, whose radius is given by the pairwise minimum of the estimated distance to the source.
This goal may be achieved via the quadratic constraint 
\begin{equation}\label{eq:range_constraint}
 \| \bbx_i - \bbx_j \|^2 \! \leq \min\{ \| \bbx_i - \bbl_i \|^2, \| \bbx_j - \bbl_j \|^2 \} \text{ for all } j \in n_i \; 
\end{equation}
which, while nonconvex, may be convexified by rearranging the right-hand side of \eqref{eq:range_constraint} and replacing the resulting maximum by the log-sum-exp function -- see \cite{Boyd2004}, Chapter 2. Thus we obtain
\begin{equation}\label{eq:range_constraint2}
 (1/2)\Big(\| \bbx_i - \bbx_j \|^2 + \log \!\left( e^{\| \bbx_i - \bbl_i \|^2} + e^{\| \bbx_j - \bbl_j \|^2} \right)\!\Big)\leq 0 \; ,
\end{equation}
%
%%%%%%%%%%%%%%%%%%%%%%%%%%%%%%%%%%%%%%%%%%%%%%%%%%%%%%%%%%%%%%%%%%%%%%%%%%%
%%%   F   I   G   U   R   E   %%%%%%%%%%%%%%%%%%%%%%%%%%%%%%%%%%%%%%%%%%%%%
%%%%%%%%%%%%%%%%%%%%%%%%%%%%%%%%%%%%%%%%%%%%%%%%%%%%%%%%%%%%%%%%%%%%%%%%%%%
%
\begin{figure*}%
\centering
\begin{subfigure}{.33\columnwidth}
\includegraphics[width=\linewidth, height = 0.75\linewidth]{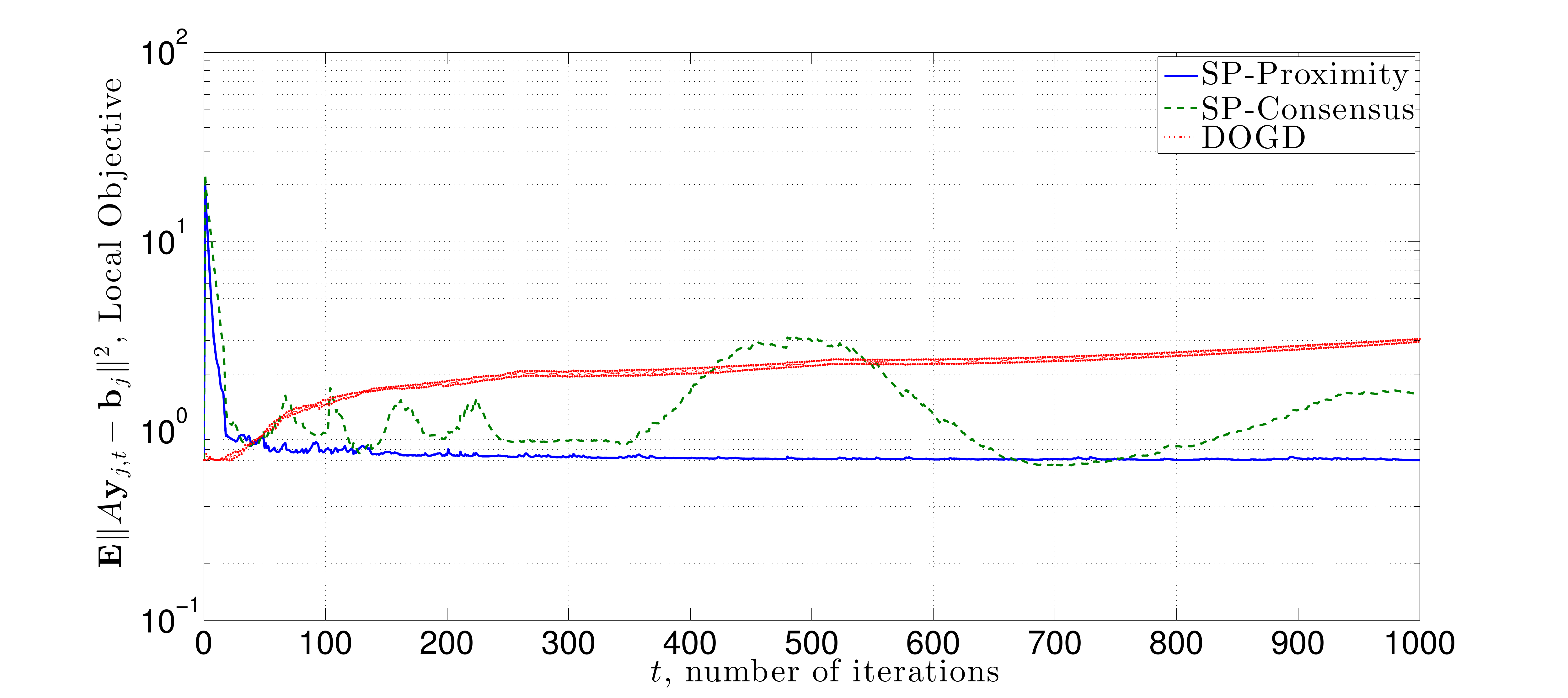}%
\caption{Local objective vs. iteration $t$}%
\label{subfiga_vary_alg}
\end{subfigure}\hfill%
\begin{subfigure}{.33\columnwidth}
\includegraphics[width=\linewidth,height = 0.75\linewidth]{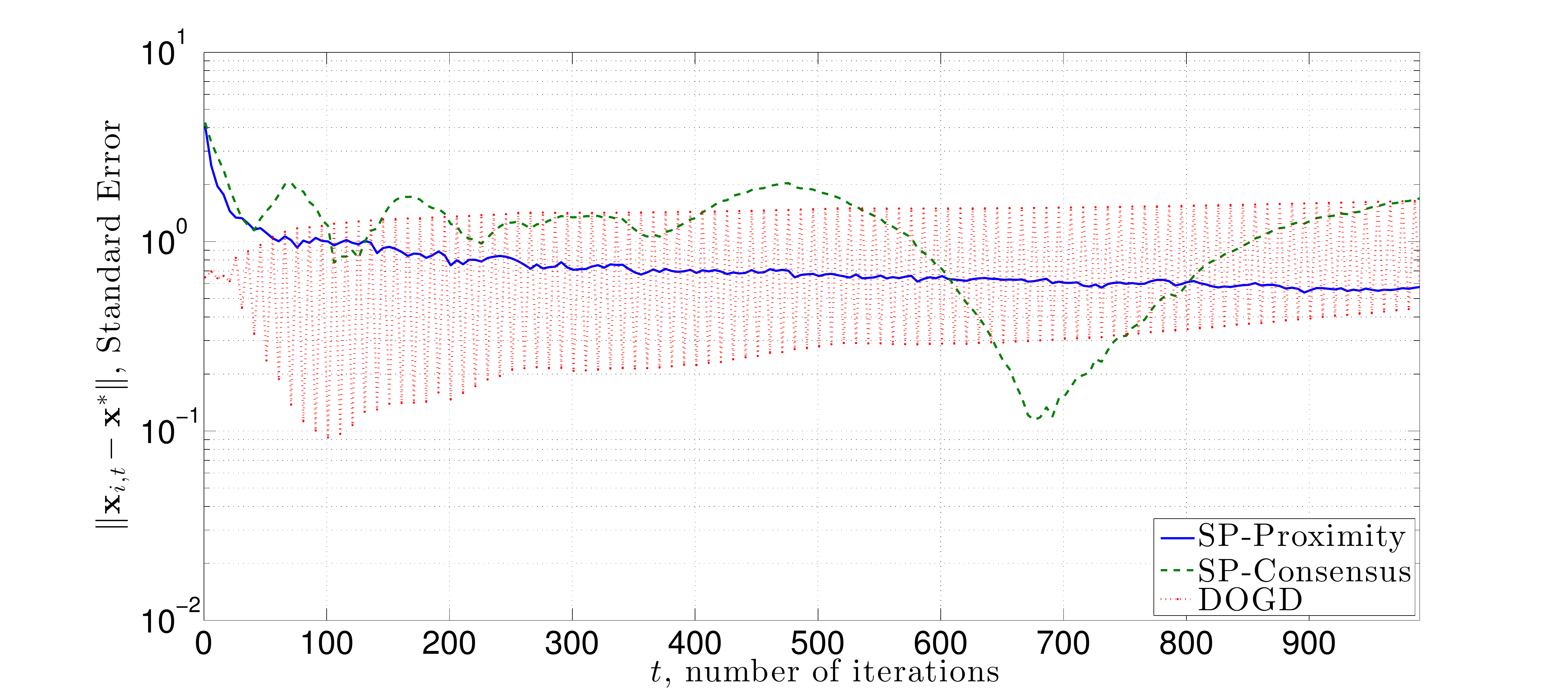}%
\caption{Standard error  $ \|  \bbx_{i,t} - \bbx^* \| $ vs. iteration $t$}%
\label{subfigb_vary_alg}
\end{subfigure}\hfill%
\begin{subfigure}{.33\columnwidth}
\includegraphics[width=\linewidth,height = 0.75\linewidth]{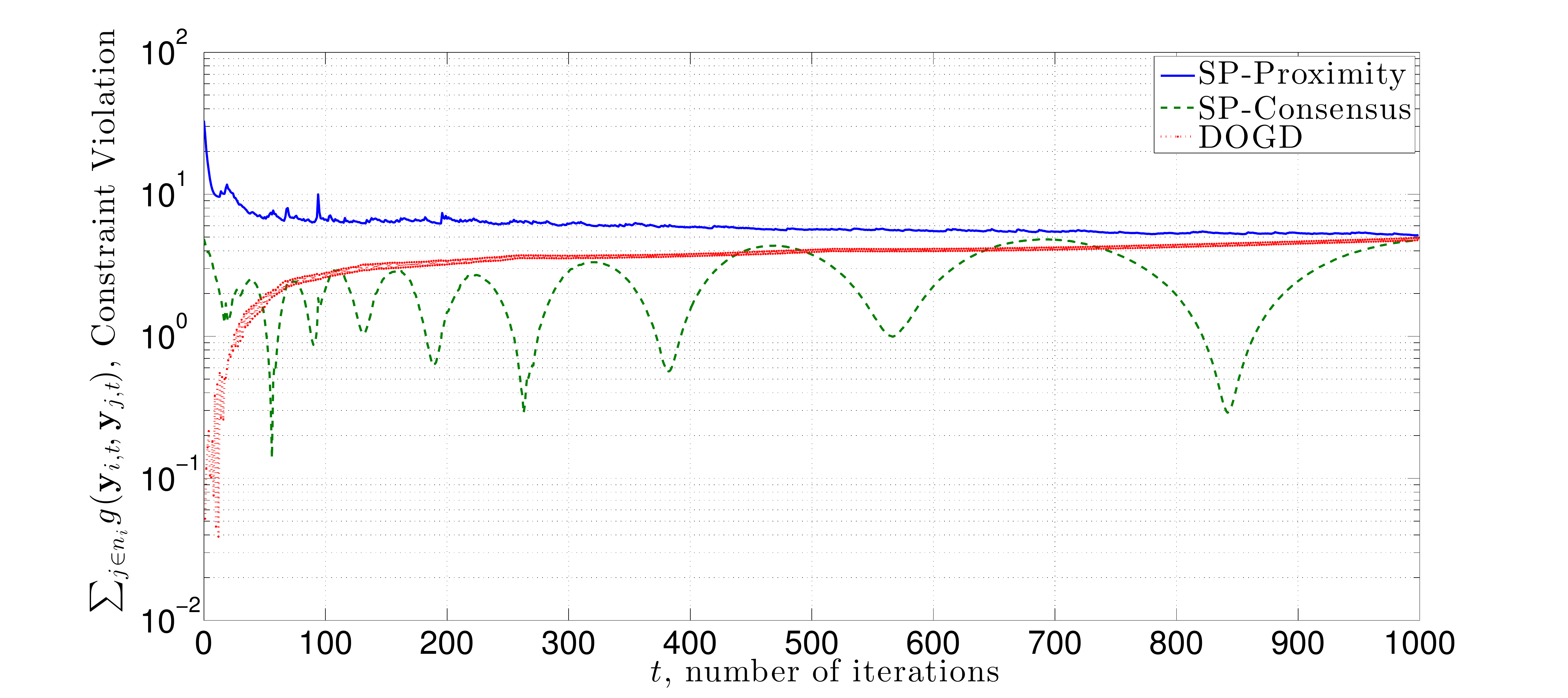}%
\caption{Constraint violation  vs. iteration $t$}%
\label{subfigc_vary_alg}%
\end{subfigure}%
\caption{Comparison of proximity and consensus algorithms on the source localization problem stated in \eqref{eq:sr_ls} using the convex approximation \eqref{eq:sr_ls_relax} for an $N=64$ node grid network deployed as an $8\times8$ square in a $1000 \times 1000$ meter region for the case that the noise perturbing observations by node $i$ is zero-mean Gaussian, with a variance proportional its distance to the source as $\sigma^2=2 \| \bbl_i - \bbx^*\|$, where $\bbl_i$ is the location of node $i$. We run the saddle point method with proximity constraints [cf \eqref{eq:localization_primal_i} - \eqref{eq:localization_dual_update}] given in \eqref{eq:sr_ls2} using dual regularization $\delta=10^{-7}$, as compared with the saddle point method which executes a consensus constraint \eqref{eq:consensus_constraint}, as well as Distributed Online Gradient Descent (DOGD) \cite{Tsianos2012}, which is a weighted averaging gradient method. For the former two, we use hybrid step-size strategy $\eps_t = \min(\eps, \eps t_0 / t)$ with $t_0=100$  and $\eps=10^{-1.5}$, and for DOGD we use constant step-size $10^{-1.5}$. We observe that the proximity-constrained saddle point method yields the best performance in terms of objective convergence and standard error, although it incurs higher levels of constraint violation.}\label{fig:vary_alg}\vspace{-3mm}
\end{figure*}
which is a convex constraint, since the later term is a composition of a monotone function with a convex function. Taking \eqref{eq:sr_ls_relax} together with the constraint \eqref{eq:range_constraint2}, and noting that a constraint on $\bbx_i$ is equivalent to a constraint on the first $p$ entries of $\bby_i$ after appending a $0$ to the $p+1$-th entry of $\bbl_i$, we may write
\begin{alignat}{2} \label{eq:sr_ls2}
&\min_{\bby\in \reals^{N(p+1)} }\ && \!\!\sum_{i=1}^N \mathbb{E}_{\bbb_i} \Big( \|\bbA_i \bby_i - \bbb_i \|^2\Big) \; ,\\
 &\text{\ s.t. } \ &&\!\! (1/2)\!\Big(\| \bby_i \!-\! \bby_j \|^2 \!+ \!\log \! \left( \! e^{\| \bby_i - \bbl_i \|^2} \!+ e^{\| \bby_j - \bbl_j \|^2}\! \right)\!\! \Big)\leq 0 , \nonumber
\end{alignat}
where the constraint for node $i$ is with respect to all of its neighbors $j\in n_i$. Observe that the problem in \eqref{eq:sr_ls2} is of the form \eqref{eq:coop_stoch_opt}. Define $g(\bby_i, \bby_j)$ as the constraint function the left-hand side of \eqref{eq:range_constraint2}. Then primal update of the saddle point method stated in \eqref{eq:sp_primal} specialized to this problem setting for node $i$ is stated as
\begin{align} \label{eq:localization_primal_i}
   \bby_{i,t+1} 
     & =  \bby_{i,t} - \eps_t  
          \Big(2\bbA_{i,t}^T\big(\bbA_{i,t}\bby_{i,t}-\bbb_{i,t}\big)
                 \qquad\quad      \\
&\quad + \sum_{j \in n_i} \bblam_{ij,t}
           \Big( \frac{e^{\|\bby_{i,t}-\bbl_i\|^2} (\bby_{i,t} -\bbl_i )}
           {e^{\|\bby_{i,t}-\bbl_i\|^2} + e^{\|\bby_{j,t} -\bbl_j\|^2}}   \!+\!  (\bby_{i,t} \!- \!\bby_{j,t})\Big), \nonumber
\end{align}
where we omit the use of set projections for simplicity, while the dual update [cf. \eqref{eq:sp_dual}] executed at the link layer of the sensor network is 
\begin{align} \label{eq:localization_dual_update}
 \bblam_{ij,t+1} =\big[(1-\delta\eps_t)  \bblam_{ij,t} + \eps_t g(\bby_{i,t}, \bby_{j,t}) \Big]_+\; .
   \end{align}

We turn to analyzing the empirical the performance of the saddle point updates \eqref{eq:localization_primal_i} - \eqref{eq:localization_dual_update} to solve localization problems in a decentralized manner, such that nodes more strongly weight the importance of sensors in closer proximity to the source in the sense of \eqref{eq:range_constraint2}. 
Besides the local objective $\mathbb{E}_{\bbb_i}  \|\bbA_i \bby_i - \bbb_i \|^2$, which we know converges to its contained optimal value, we also study the standard error to the source signal $\bbx^*$, denoted as $\| \bbx_{i,t} - \bbx^*\|$. Recall that we recover $\bbx_{i,t}$ from $\bby_{i,t}$ by taking its first $p$ elements. We further consider the magnitude of the constraint violation for this problem, which when considering the proximity constrained problem in \eqref{eq:sr_ls2}, is given by
\begin{align}\label{eq:constraint_violation_prox}
\sum_{j\in n_i } (1/2) g(\bby_{i,t}, \bby_{j,t}) &=\sum_{j\in n_i } (1/2)\!\Big(\| \bby_{i,t} \!-\! \bby_{j,t} \|^2 \!  \\
&\quad + \!\log \! \left( \! e^{\| \bby_{i,t} - \bbl_{i,t} \|^2} \!+ e^{\| \bby_{j,t} - \bbl_j \|^2}\! \right)\!\! \Big)\nonumber\;,
\end{align}
and when implementing consensus methods, is given by 
\begin{align}\label{eq:constraint_violation_consensus}
\sum_{j\in n_i }  h(\bby_{i,t}, \bby_{j,t}) = \sum_{j\in n_i } \| \bby_{i,t} -  \bby_{j,t} \|
\end{align}
for a randomly chosen sensor in the network. 

Throughout the rest of this section, we fix the dual regularization parameter $\delta=10^{-7}$, and  study the performance of the saddle point method with proximity constraints as compared with two methods which attempt to satisfy consensus constraints. We further analyze the saddle point method in \eqref{eq:localization_primal_i} - \eqref{eq:localization_dual_update} for a variety of network sizes to understand the practical effect of the learning rate on the number of sensors, and for different spatial deployment strategies which induce different network topologies. 

\subsection{Consensus Comparison}\label{subsec:consensus}

In this subsection, we compare the saddle point method on a proximity constrained problem as compared with methods which implement variations of the consensus protocol. In particular, we run the saddle point method for the localization problem given in \eqref{eq:localization_primal_i} - \eqref{eq:localization_dual_update} with proximity constraints, as compared with the same primal-dual scheme when the consensus constraint in \eqref{eq:consensus_constraint} is used. We further compare these instantiations of the saddle point method with distributed online gradient descent (DOGD) \cite{Tsianos2012}, which is a scheme that operates by having each node selecting its next iterate by taking a weighted average of its neighbors and descending through the negative of the local stochastic gradient. For each of these methods, we run the localization procedure for a total of $1000$ iterations for $\tilde{T}=100$ different runs when each node initializes its local variable $\bby_{i,0}$ uniformly at random from the unit interval, and plot the sample mean of the results.

We consider problem instances of \eqref{eq:sr_ls} when the number $N=64$ of sensors is fixed, and are spatially deployed in a grid formation as a $8\times8$ square in a planar ($p=2$) region of size $1000\times1000$. 
Moreover, the noise perturbing the observations at node $i$ is zero-mean Gaussian, with a variance proportional its distance to the source as $\sigma^2=2 \| \bbl_i - \bbx^*\|$, where $\bbl_i$ is the location of node $i$, and the true source signal $\bbx^*$ is located at the average location of the sensors. For the saddle point methods, we find a hybrid step-size strategy to be most effective, and hence set  $\eps_t = \min(\eps, \eps t_0 / t)$ with $t_0=100$  and $\eps=10^{-1.5}$. For DOGD, we find best performance to correspond to using a constant outer step-size $\eps=10^{-1.5}$, along with a halving scheme step-size in the inner recursive averaging loop \cite{Tsianos2012}.
%%%%%%%%%%%%%%%%%%%%%%%%%%%%%%%%%%%%%%%%%%%%%%%%%%%%%%%%%%%%%%%%%%%%%%%%%%%
%%%   F   I   G   U   R   E   %%%%%%%%%%%%%%%%%%%%%%%%%%%%%%%%%%%%%%%%%%%%%
%%%%%%%%%%%%%%%%%%%%%%%%%%%%%%%%%%%%%%%%%%%%%%%%%%%%%%%%%%%%%%%%%%%%%%%%%%%
%
\begin{figure*}%
\centering
\begin{subfigure}{.33\columnwidth}
\includegraphics[width=1\linewidth, height = 0.75\linewidth]{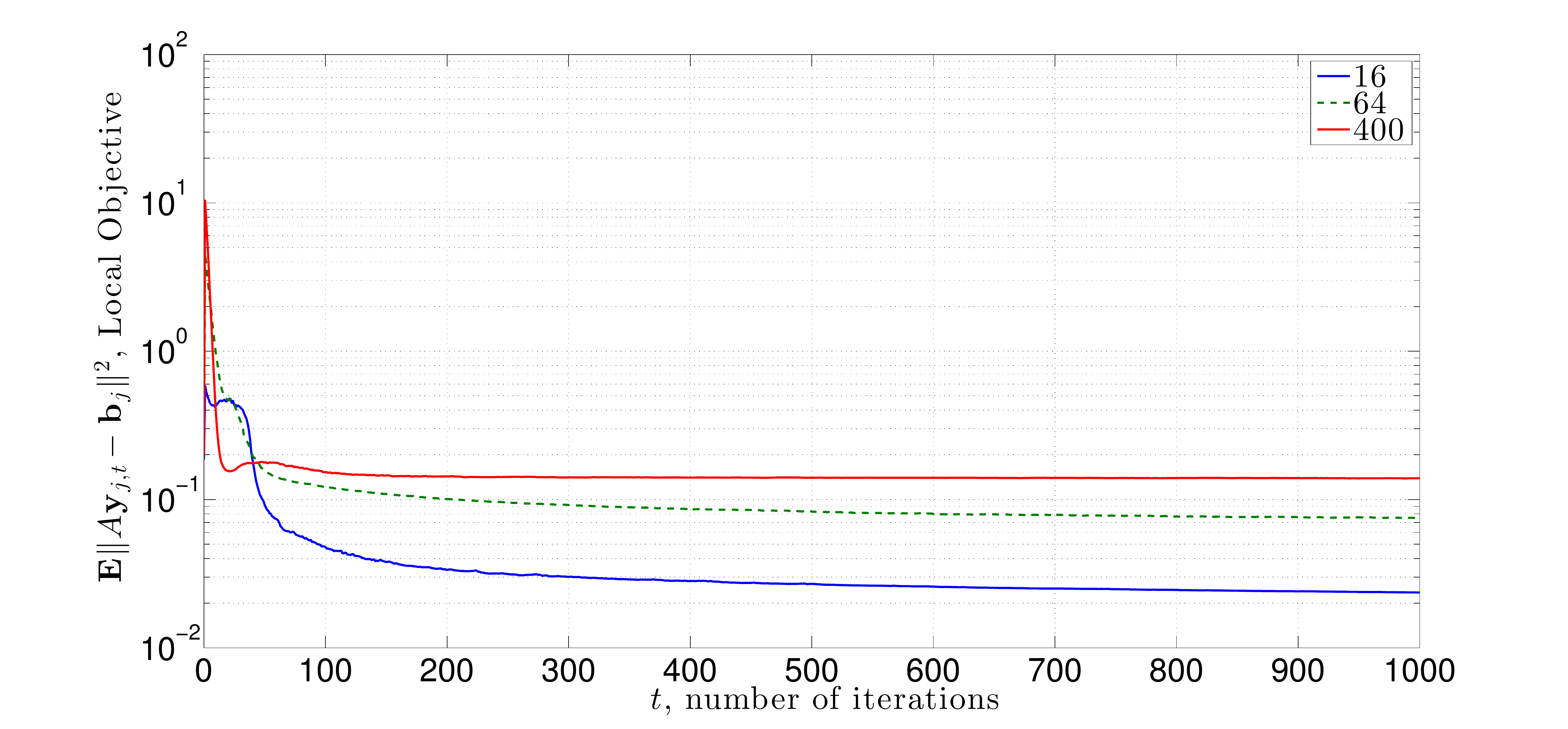}%
\caption{Local objective vs. iteration $t$}%
\label{subfiga_vary_n}%
\end{subfigure}\hfill%
\begin{subfigure}{.33\columnwidth}
\includegraphics[width=1\linewidth,height = 0.75\linewidth]{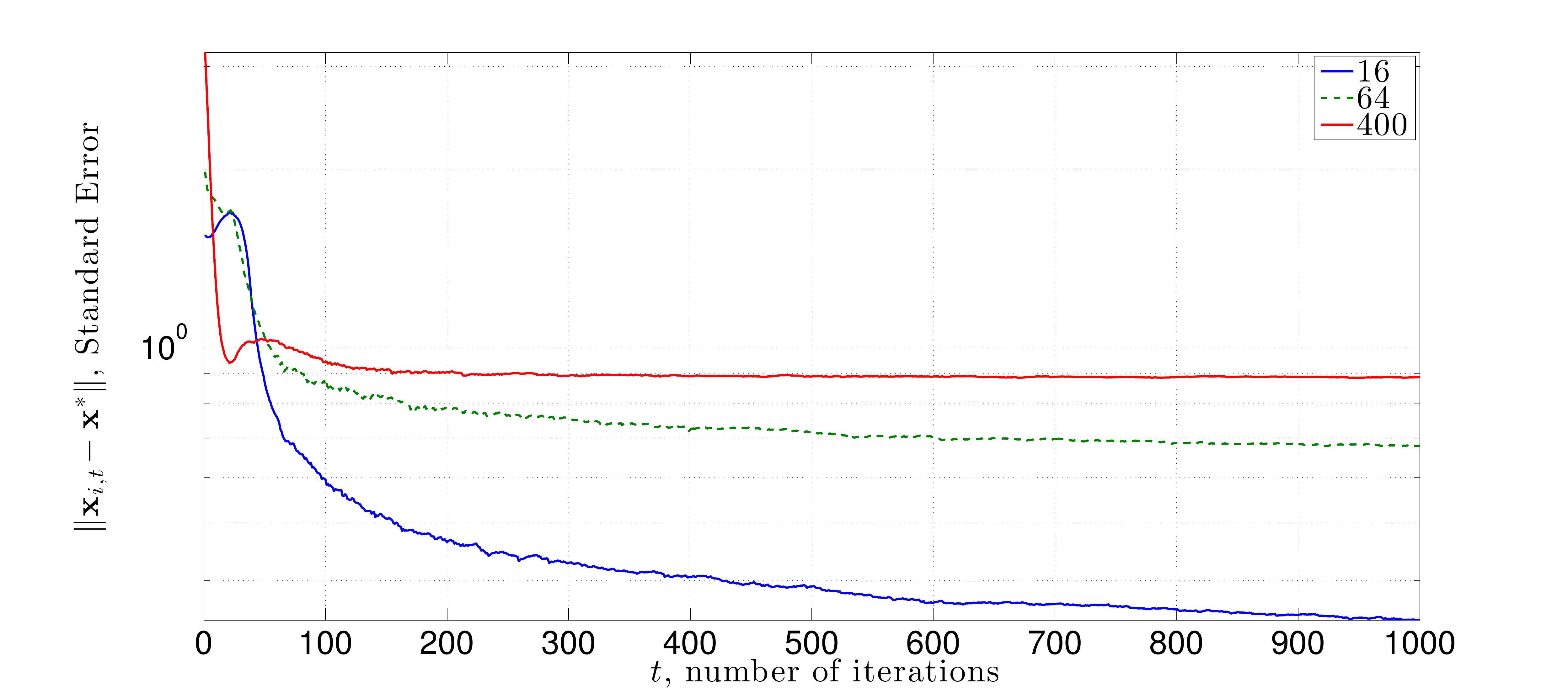}%
\caption{Standard error  $ \|  \bbx_{i,t} - \bbx^* \| $ vs. iteration $t$}%
\label{subfigb_vary_n}%
\end{subfigure}\hfill%
\begin{subfigure}{.33\columnwidth}
\includegraphics[width=1\linewidth,height = 0.75\linewidth]{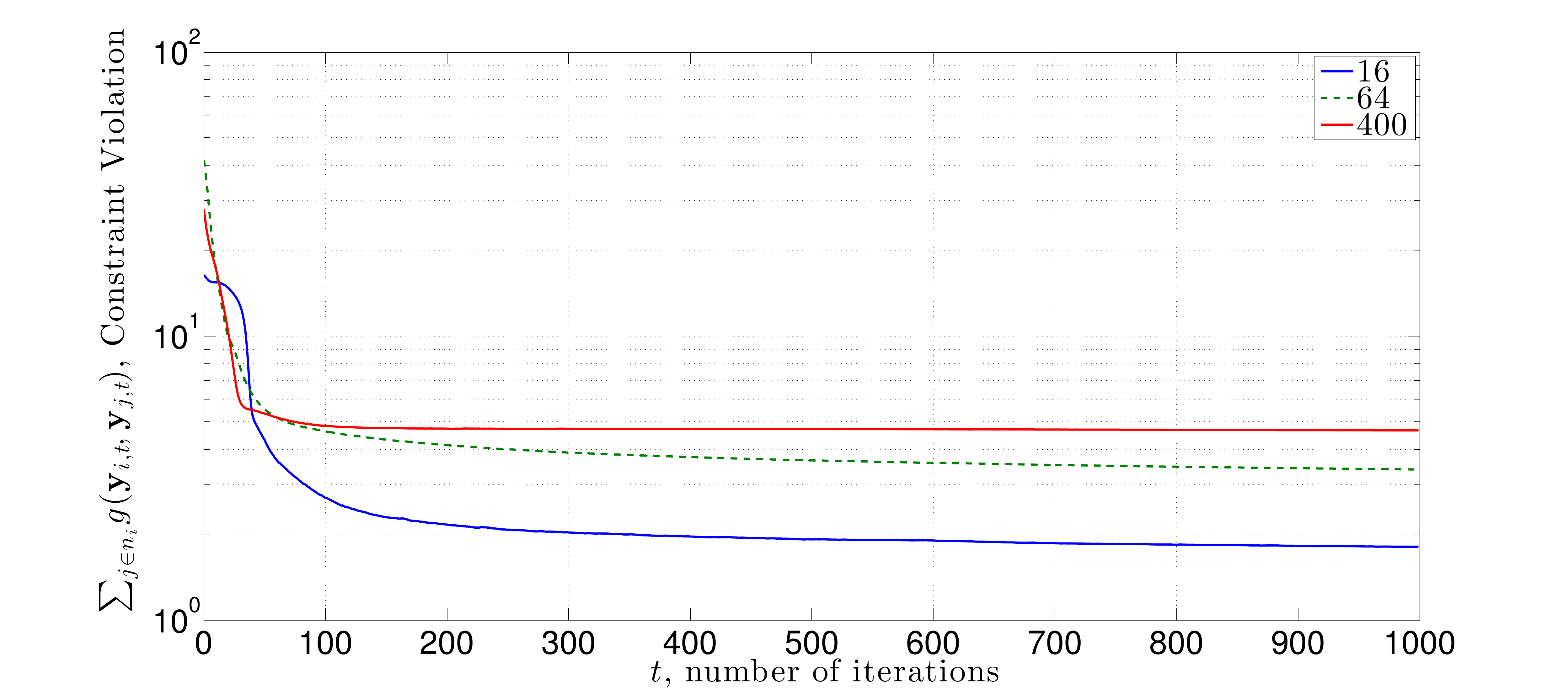}%
\caption{Constraint violation  vs. iteration $t$}%
\label{subfigc_vary_n}%
\end{subfigure}%
\caption{Comparison of the saddle point method with proximity constraints [cf. \eqref{eq:localization_primal_i} - \eqref{eq:localization_dual_update}] with dual regularization $\delta=10^{-7}$ and hybrid step-size strategy $\eps_t = \min(\eps, \eps t_0 / t)$ with $t_0=100$  and $\eps=10^{-1.5}$ on the source localization problem stated in \eqref{eq:sr_ls} using the convex approximation \eqref{eq:sr_ls_relax}. We fix the network topology as a grid and vary the number of sensors as $N=16$, $N=64$, and $N=400$ which are deployed in a square region of size $1000 \times 1000$ square meters. The noise perturbing observations by sensor $i$ is zero-mean Gaussian, with a variance proportional its distance to the source as $\sigma^2=0.5 \| \bbl_i - \bbx^*\|$, where $\bbl_i$ is the location of node $i$. Observe that in larger networks, the rate at which nodes are able to localize the source is slower in terms of objective convergence and standard error to the true source. Moreover, we see that the level of constraint violation is larger with increasing $N$. }\label{fig:vary_n}\vspace{-3mm}
\end{figure*}

We plot the results of these methods for this problem instance in Figure \ref{fig:vary_alg} for an arbitrarily chosen node $i \in V$. Observe that the saddle point method which implements network proximity constraints method yields the best performance in terms of objective convergence. In particular, by $t=500$ iterations, in Figure \ref{subfiga_vary_alg} we observe that the saddle point algorithm implemented with proximity constraints (SP-Proximity) achieves objective convergence to a neighborhood, i.e. $\mathbb{E}_{\bbb_i} \|\bbA_i \bby_{i,t} - \bbb_i \|^2 \leq 1$. In contrast, the saddle point with consensus constraints (SP-Consensus) and DOGD respectively experience numerical oscillations and divergent behavior after a burn-in period of $t=100$.

This trend is confirmed in the plot of the standard error to the optimizer $\|\bbx_{i,t} - \bbx^*\|$ of the original problem \eqref{eq:sr_ls} in Figure \ref{subfigb_vary_alg}. We see that SP-Proximity yields convergence to a neighborhood between $10^{-1}$ and $1$ by $t=200$ iterations, whereas SP-Consensus and DOGD experience numerical oscillations and do not appear to localize the source signal $\bbx^*$.
While SP-Proximity exhibits superior behavior in terms of objective and standard error convergence, it incurs larger levels of constraint violation than its consensus counterpoints, as may be observed in Figure \ref{subfigc_vary_alg}. To be specific, SP-Proximity on average experiences constraint violation [cf. \eqref{eq:constraint_violation_prox}] on average an order of magnitude larger than SP-Consensus and DOGD [cf. \eqref{eq:constraint_violation_consensus}]  for the first $t=400$ iterations. After this benchmark, the magnitude of the constraint of the different methods converges to around $5$. Thus, we see that achieving smaller constraint violation and implementing consensus constraints may lead to inferior source localization accuracy.

\subsection{Impact of Network Size}\label{subsec:size}

In this subsection, we study the effect of the size of the deployed sensor network on the ability of the proximity-constrained saddle point method to effectively localize the source signal. We fix the topology of the deployed sensors as a grid network, and again set the source signal $\bbx^*$ to be the average of node positions in a planar ($p=2$) spatial region $\ccalA$ of size $1000\times 1000$ meters. We set the noise distribution which perturbs the range measurements of node $i$ to be zero-mean Gaussian with variance $.5 \| \bbl_i - \bbx^*\|$. We run the algorithm stated in \eqref{eq:localization_primal_i} - \eqref{eq:localization_dual_update}  with hybrid step-size strategy $\eps_t = \min(\eps, \eps t_0 / t)$ with $t_0=100$  and $\eps=10^{-1.5}$ for a total of $T=1000$ iterations for $\tilde{T}=100$ total runs when each node initializes its local variable $\bby_{i,0}$ uniformly at random from the unit interval, and plot the sample mean results for problem instances of \eqref{eq:sr_ls2} when the network size is varied as $N=16$, $N=64$, $N=400$, which correspond respectively to $4\times4$, $8\times 8$, and $20\times20$ grid sensor formations.

We plot the results of this numerical setup in Figure \ref{fig:vary_n} for a randomly chosen sensor in the network. Observe that in Figure \ref{subfiga_vary_n}, which shows the convergence behavior in terms of the local objective $\mathbb{E}_{\bbb_i} \|\bbA_i \bby_{i,t} - \bbb_i \|^2$ versus iteration $t$, that the rate at which sensors are able to localize the source is comparable across the different network sizes; however, the convergence accuracy is higher in smaller networks. In particular, by $t=1000$, we have observe the objective converges to respective values $0.03$, $0.08$, and $0.14$ for the $N=16$, $N=64$, $N=400$ node networks. This relationship between convergence accuracy  and number of sensors in the network is corroborated in the plot of the standard error  $\|\bbx_{i,t} - \bbx^*\|$ to the true source $\bbx^*$ in Figure \ref{subfigb_vary_n}. We see that the standard error across the different networks converges to within a radius of $1$ to the optimum, but the rate at which convergence is exhibited is comparable across the different network sizes. In particular, by $t=400$, we observe the standard error benchmarks $0.41$, $0.74$, and $0.9$ for the $N=16$, $N=64$, and $N=400$ node networks.

A similar pattern may be gleaned from Figure \ref{subfigc_vary_n}, in which we plot the magnitude of the constraint violation $\sum_{j\in n_i} g(\bby_{i,t}, \bby_{j,t})$ as given in \eqref{eq:constraint_violation_prox} with iteration $t$. Observe that for the networks with $N=16$, $N=64$, and $N=400$ sensors, respectively, we have the constraint violation benchmarks $2.1$, $4$, and $4.74$ by $t=300$. Moreover, the rate at which benchmarks are achieved is comparable across the different network sizes, such that the primary difference in the dual domain is the asymptotic magnitude of constraint violation, but not dual variable convergence rate.
%%%%%%%%%%%%%%%%%%%%%%%%%%%%%%%%%%%%%%%%%%%%%%%%%%%%%%%%%%%%%%%%%%%%%%%%%%%
%%%   F   I   G   U   R   E   %%%%%%%%%%%%%%%%%%%%%%%%%%%%%%%%%%%%%%%%%%%%%
%%%%%%%%%%%%%%%%%%%%%%%%%%%%%%%%%%%%%%%%%%%%%%%%%%%%%%%%%%%%%%%%%%%%%%%%%%%
%
\begin{figure*}%
\centering
\begin{subfigure}{.33\columnwidth}
\includegraphics[width=\linewidth, height = 0.75\linewidth]{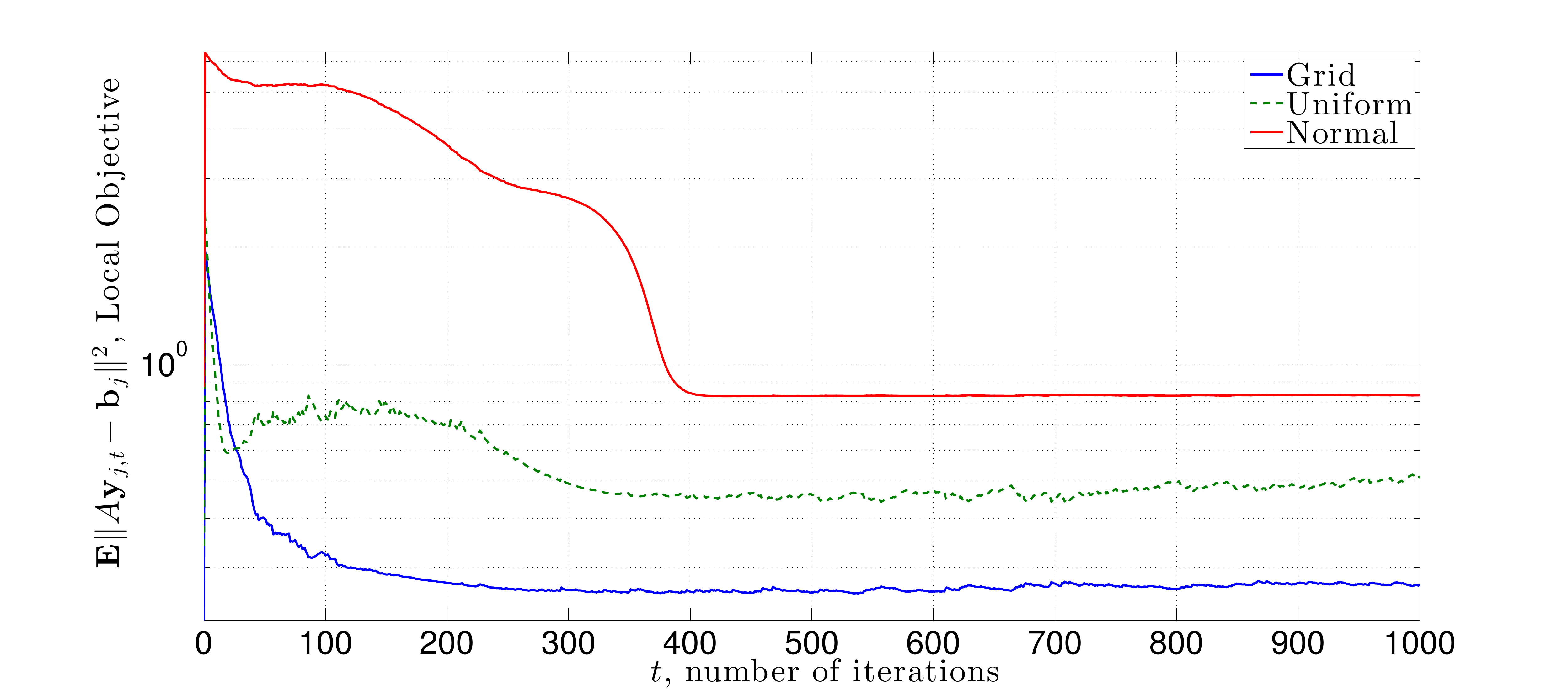}%
\caption{Local objective vs. iteration $t$}%
\label{subfiga_vary_network}%
\end{subfigure}\hfill%
\begin{subfigure}{.33\columnwidth}
\includegraphics[width=\linewidth,height = 0.75\linewidth]{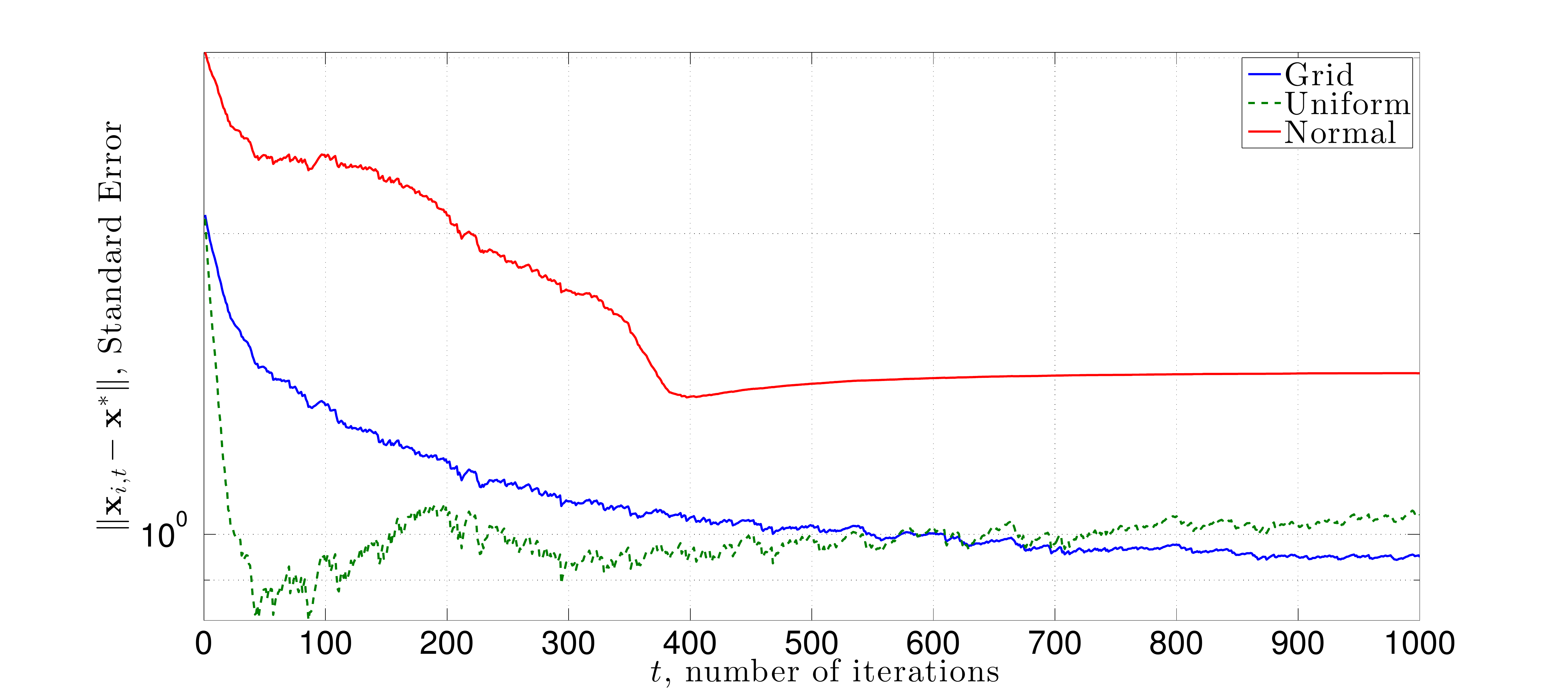}%
\caption{Standard error  $ \|  \bbx_{i,t} - \bbx^* \| $ vs. iteration $t$}%
\label{subfigb_vary_network}%
\end{subfigure}\hfill%
\begin{subfigure}{.33\columnwidth}
\includegraphics[width=\linewidth,height = 0.75\linewidth]{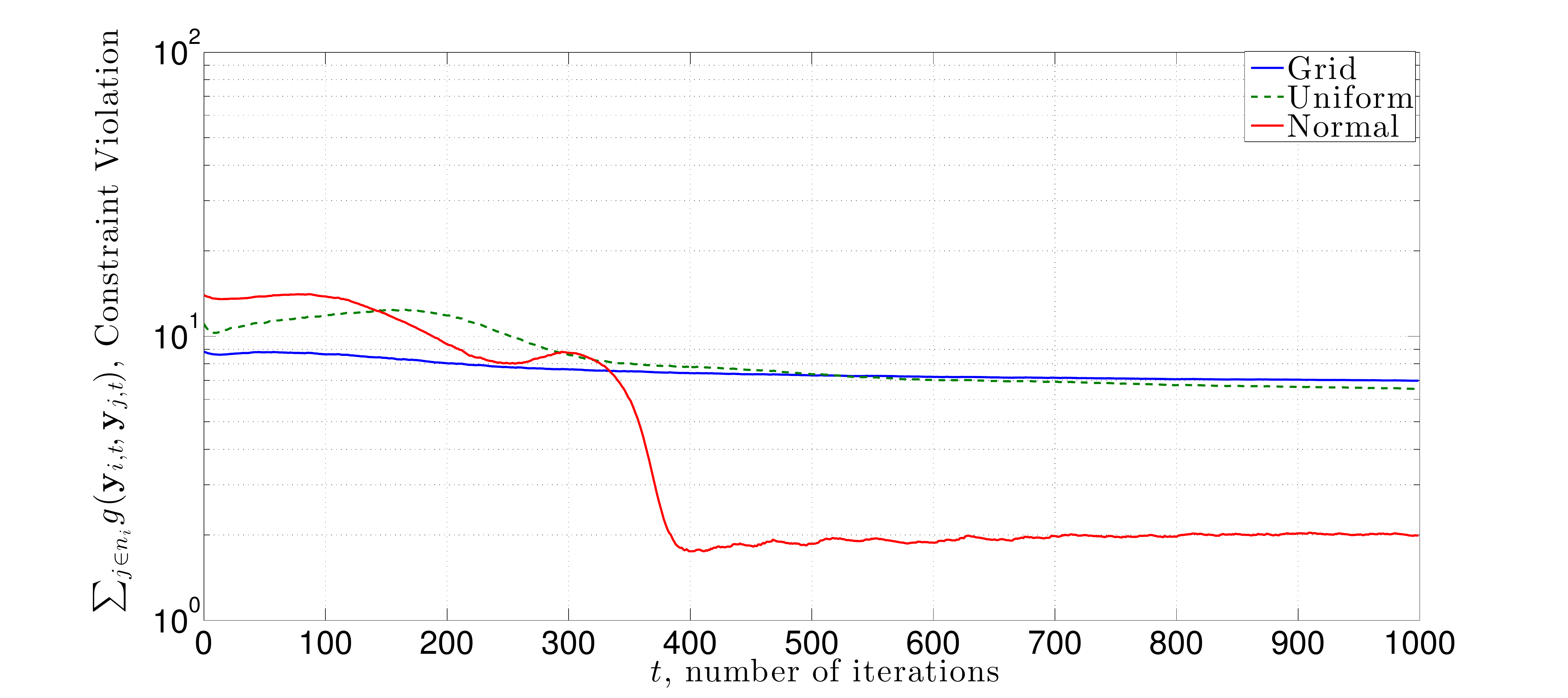}%
\caption{Constraint violation  vs. iteration $t$}%
\label{subfigc_vary_network}%
\end{subfigure}%
\caption{Numerical results on the localization problem stated in \eqref{eq:sr_ls} using the convex approximation \eqref{eq:sr_ls_relax} for the saddle point method with proximity constraints [cf. \eqref{eq:localization_primal_i} - \eqref{eq:localization_dual_update}] with hybrid step-size strategy $\eps_t = \min(\eps, \eps t_0 / t)$ with $t_0=100$  and $\eps=10^{-1.5}$. We run the algorithm for a variety of spatial deployment strategies, which induce different network topologies. We consider a square region of size $1000 \times 1000$ square meters, and deploy nodes in grid formations, uniformly at random, and according to a two-dimensional Normal distribution. In the later two cases, sensors which are a distance of $50$ meters are less are connected by an edge. The noise perturbing observations by sensor $i$ is zero-mean Gaussian, with a variance proportional its distance to the source as $\sigma^2=0.5 \| \bbl_i - \bbx^*\|$, where $\bbl_i$ is the location of node $i$. We see that while the Normal configuration yields the worst localization performance, it achieves the least levels of constraint violation. In contrast, Uniform and Grid configurations both are effective spatial deployment strategies to localize the source in terms of local objective convergence and standard error. }\label{fig:vary_network}\vspace{-3mm}
\end{figure*}
\subsection{Effect of Spatial Deployment}\label{subsec:topology}

We turn to studying the impact of the way in which sensors are spatially deployed on their ability to localize the source signal, which implicitly is an analysis of the impact of the network topology on the empirical convergence behavior. To do so, we consider a  problem instance in which the source signal $\bbx^*$ is located at the average of sensor positions in the network in a planar ($p=2$) spatial region $\ccalA$ of size $1000\times 1000$ meters. Moreover, the noise distribution which perturbs the range measurements received at node $i$ is fixed as zero-mean Gaussian with variance $.5 \| \bbl_i - \bbx^*\|$, implying that nodes which are closer to the source receive observations with higher SNR. Each node initializes its local variable $\bby_{i,0}$ uniformly at random from the unit interval, and then executes the saddle point method stated in \eqref{eq:localization_primal_i} - \eqref{eq:localization_dual_update} for a total of $T=1000$ iterations for $\tilde{T}=100$ total runs. We consider the sample mean results of the $\tilde{T}=100$ for problem instances of \eqref{eq:sr_ls2} when the sensor deployment strategy is either a grid formation, uniformly at random, or according to a two-dimensional Gaussian distribution. In the later two cases, sensors which are closer than a distance of $50$ meters are connected. Since in general random networks of these types will not be connected, we repeatedly generate such networks until we obtain the first one which has the a comparable Fiedler number (second-smallest eigenvalue of the graph Laplacian matrix) as the grid network, which is a standard measure of network connectivity -- see \cite{Chung97a}, Ch. $1$ for details. 

We display the results of this localization experiment in Figure \ref{fig:vary_network}. In Figure \ref{subfiga_vary_network}, we plot the local objective as compared with iteration $t$ across these different sensor deployment strategies. We see that sensor localization performance is best in terms of objective convergence in the grid network, followed by network topologies generated from uniform and Normal spatial deployment strategies. In particular, by $t=400$, the grid, Uniform, and Normal sensor networks achieve the objective ($\mathbb{E}_{\bbb_i} \|\bbA_i \bby_{i,t} - \bbb_i \|^2$) benchmarks $0.26$, $0.45$, and $0.83$. This trend is complicated by our analysis of these sensor networks' ability to learn the true source $\bbx^*$ as measured by the standard error  $\|\bbx_{i,t} - \bbx^*\|$ versus iteration $t$ in Figure \ref{subfigb_vary_network}. In particular, to achieve the benchmark $\|\bbx_{i,t} - \bbx^*\| \leq 1$ we see that the Uniform topology requires $t=26$ iterations, whereas the grid network requires $t=557$ iterations, and the Normal network does not achieve error bound by $t=1000$. However, we observe that the grid network experiences more stable convergence behavior in terms of its error sequence, as compared with the other two networks.

In Figure \ref{subfigc_vary_network}, we display the constraint violation [cf. \eqref{eq:constraint_violation_prox}] incurred by the proximity-constrained saddle point method when we vary the sensor deployment strategy. Observe that the grid and Uniform network topologies incur comparable levels of constraint violation, whereas the sensor network induced by choosing spatial locations according to a two-dimensional Gaussian distribution is able to maintain closer levels of network proximity by nearly an order of magnitude. %Thus we see that ensuring that the estimates at distinct sensors are close indeed may harm estimation performance in terms of objective and standard error convergence.

%!TEX root = root.tex
\section{Conclusion} \label{sec:conclusion}
We developed multi-agent stochastic optimization problems where the hypothesis that all agents are trying to learn common parameters may be violated. In doing so, agents to make decisions which give preference to locally observed information while incorporating the relevant information of others. This problem class incorporates sequential estimation problems in multi-agent settings where observations are independent but \emph{not identically} distributed. We formulated this task as a decentralized stochastic program with convex proximity constraints which incentivize distinct nodes to make decisions which are close to one another. We considered an augmented Lagrangian relaxation of the problem, to which we apply a stochastic variant of the saddle point method of Arrow and Hurwicz to solve it. We established that under a constant step-size regime the time-average suboptimality and constraint violation are contained in a neighborhood whose radius vanishes with increasing number of iterations (Theorem \ref{theorem1}). As a consequence, we obtain in Corollary \ref{corollary2} that the average primal vectors converge to the optimum while satisfying the network proximity constraints.

Numerical analysis on a random field estimation problem in a sensor network illustrated the benefits of using the saddle point method with proximity constraints as compared with a simple LMMSE estimator scheme. We find that these benefits are more pronounced in problem instances with lower SNR and larger spatial regions in which sensors are deployed. We further considered a source localization problem in a sensor network, where sensors collect noisy range estimates whose SNR is proportional to their distance to the true source signal. In this problem setting, the proximity-constrained saddle point method outperforms methods which attempt to execute consensus constraints.

%!TEX root = root.tex

\section*{Appendix A: Proof of Proposition \ref{prop1}} \label{prop1_proof}

Begin by observing that since we assume that $h_i(\bbx_i, \bbx_{j} ) = h_j(\bbx_j, \bbx_{i} )$ for all $\bbx_i\in\ccalX$ and $\bbx_{j}\in\ccalX$  it must be that gradients are also equal and that, in particular,
\begin{align}  \label{eq:dict_grads_are_equal}
   \nabla_{\bbx_i} h_i(\bbx_{i,t}, \bbx_{j,t}) = 
   \nabla_{\bbx_i} h_j(\bbx_{j,t}, \bbx_{i,t}) .
\end{align}
To compute the primal stochastic gradient of the Lagrangian in \eqref{eq:lagrangian}, observe that in the instantaneous Lagrangian in \eqref{eq:stoch_lagrangian} only a few summands depend on $\bbx_i$. In the first sum only the one associated with the local objective $f_i(\bbx_i,\bbtheta_{i,t})$ depends on $\bbx_i$. In the second sum the terms that depend on $\bbx_i$ include the local constraints $h_i(\bbx_i,\bbx_j ) - \gamma_{ij}$ and the neighboring constraints $h_j(\bbx_j,\bbx_i ) - \gamma_{ji}$. Taking gradients of these terms and recalling the equality in \eqref{eq:dict_grads_are_equal} yields,
\begin{align}  \label{eq:dict_grad_appendix}
 \nabla_{\bbx_i} \hat{\ccalL}_t (\bbx_t, \bblam_t) =\ & 
 	 \nabla_{\bbx_i} f_i (\bbx_{i,t} ; \bbtheta_{i,t} )   	  \\ \nonumber & 
	  + \sum_{j \in n_i} (\bblam_{ij,t} +  \bblam_{ji,t})^T  \nabla_{\bbx_i} h_i(\bbx_{i,t}, \bbx_{j,t}) .
 \end{align}
Writing \eqref{eq:sp_primal} componentwise and substituting $\nabla_{\bbx_i} \hat{\ccalL}_t (\bbx_t, \bblam_t)$ for its expression in \eqref{eq:dict_grad_appendix}, the result in \eqref{eq:local_primal_update} follows.

%To develop the dual variable update in \eqref{eq:local_dual_update}, note that the dual feasible set $\reals_+^M$ is a Cartesian product of individual feasible sets $\reals_+$. Thus, it follows that we can rewrite \eqref{eq:sp_dual} component-wise as
%%
%\begin{equation} \label{eq:projection_decomposed}
%    \lam_{ij,t+1} =  \Big[(1-\eps_t \delta)\lam_{ij,t} + \eps_t \nabla_{\lam_{ij}} \hat{\ccalL} (\bbx_{t+1}, \bblam_t) \Big]_+ .
%\end{equation}
%%
To prove \eqref{eq:local_dual_update} we just need to compute the gradient $\hat{\ccalL}_t (\bbx_{t}, \bblam_t)$ of the stochastic Lagrangian with respect to the Lagrange multipliers associated with edge $(i,j)$. By noting that only one summand in \eqref{eq:stoch_lagrangian} depends on this multiplier we conclude that
\begin{equation} \label{eq:dual_grad_lam_appendix}
   \nabla_{\lam_{ij}} \hat{\ccalL}_t  (\bbx_{t}, \bblam_t)
      =   h_i  (\bbx_{i,t}, \bbx_{j,t}) - \gamma_{ij} - \eps_t \delta \lambda_{ij,t}\; .
\end{equation}
After gathering terms in \eqref{eq:dual_grad_lam_appendix} and substituting the result into \eqref{eq:sp_dual}, we obtain \eqref{eq:local_dual_update}. $\qed$

\section*{Appendix B: Proof of Lemma \ref{lemma1}} \label{lemma2_proof}
Consider the squared 2-norm of the difference between the iterate $\bbx_{t+1}$ at time $t+1$ and an arbitrary feasible point $\bbx\in \ccalX$ and use \eqref{eq:sp_primal} to express $\bbx_{t+1}$ in terms of  $\bbx_{t}$,
\begin{align} \label{eq:primal_dist} 
   \|\bbx_{t+1}-\bbx \|^2 
       = \|\ccalP_{\ccalX^N}[\bbx_t - \eps_t \nabla_\bbx \hat{\ccalL}_t(\bbx_t,\bblam_t)]-\bbx \|^2.
\end{align}
Since $\bbx\in X$ the distance between the projected vector $\ccalP_X[\bbx_t - \eps \nabla_\bbx \hat{\ccalL}_t(\bbx_t,\bblam_t)]$ and $\bbx$ is smaller than the distance before projection. Use this fact in \eqref{eq:primal_dist} and expand the square to write
\begin{align} \label{eq:primal_sq_expand} 
   \|\bbx_{t+1}-\bbx \|^2 \
      \leq\ & \|\bbx_t-\eps_t\nabla_\bbx\hat{\ccalL}_t(\bbx_t,\bblam_t)-\bbx\|^2  \nonumber\\ 
      =   \ & \|\bbx_t - \bbx\|^2 
              - 2\eps_t\nabla_\bbx\hat{\ccalL}_t(\bbx_t,\lambda_t)^T(\bbx_t-\bbx) \nonumber\\&\quad
              + \eps_t^2\|\nabla_\bbx\hat{\ccalL}_t(\bbx_t,\bblam_t)\|^2.
\end{align}
We reorder terms of the above expression such that the gradient inner product is on the left-hand side, yielding
\begin{align} \label{eq:primal_grad_neq}
   \nabla_\bbx&\hat{\ccalL}_t(\bbx_t,\bblam_t)^T(\bbx_t-\bbx)  \\\nonumber
      & \leq   \frac{1}{2\eps_t}\left(\|\bbx_t-\bbx\|^2 \!-\! \|\bbx_{t+1}-\bbx\|^2\right) 
             + \frac{\eps_t}{2}\|\nabla_\bbx\hat{\ccalL}_t(\bbx_t,\bblam_t)\|^2.
\end{align}
Observe now that since the functions  $f_{i,t}(\bbx_i,\bbtheta)$ and $h_{ij}(\bbx_i,\bbx_j)$ are convex, the online Lagrangian is a convex function of $\bbx$ [cf. \eqref{eq:lagrangian}]. Thus, it follows from the first order convexity condition that  
\begin{equation} \label{eq:primal_cvx}
   \hat{\ccalL}_t(\bbx_t,\bblam_t) - \hat{\ccalL}_t(\bbx,\bblam_t) 
      \leq \nabla_\bbx\hat{\ccalL}_t(\bbx_t,\bblam_t)^T(\bbx_t-\bbx).
\end{equation}
Substituting the upper bound in \eqref{eq:primal_grad_neq} for the right hand side of the inequality in \eqref{eq:primal_cvx} yields
\begin{align} \label{eq:lagrange_primal_neq}
   \hat{\ccalL}_t&(\bbx_t,\bblam_t) - \hat{\ccalL}_t(\bbx,\bblam_t)   \\\nonumber
      & \leq   \frac{1}{2\eps}\left(\|\bbx_t-\bbx\|^2-\|\bbx_{t+1}-\bbx\|^2\right) 
             + \frac{\eps_t}{2}\|\nabla_\bbx\hat{\ccalL}_t(\bbx_t,\bblam_t)\|^2.
\end{align}
We set this analysis aside and proceed to repeat the steps in \eqref{eq:primal_dist}-\eqref{eq:lagrange_primal_neq} for the distance between the iterate $\bblam_{t+1}$ at time $t+1$ and an arbitrary multiplier $\bblam$.
\begin{align} \label{eq:dual_dist} 
   \|\bblam_{t+1}-\bblam \|^2  
   	=\|  [ \bblam_t + \eps_t \nabla_{\bblam} \hat{\ccalL}_t(\bbx_t,\bblam_t) ]_+-\bblam \|^2,
\end{align}
where we have substituted \eqref{eq:sp_dual} to express $\bblam_{t+1}$ in terms of $\bblam_t$.
Using the non-expansive property of the projection operator in \eqref{eq:dual_dist} and expanding the square, we obtain
\begin{align} \label{eq:dual_sq_expand}
   \| \bblam_{t+1}-\bblam\|^2 
   	&\leq \|\bblam_t + \eps_t \nabla_{\bblam}\hat{\ccalL}_t(\bbx_t,\bblam_t) -  \bblam \|^2.  \\
	 & = \|\bblam_t - \bblam\|^2 
  	+2\eps \nabla_{\bblam}\hat{\ccalL}_t(\bbx_t,\lam_t)^T(\bblam_t-\bblam) \nonumber\\
  	&\quad+ \eps^2 \| \nabla_{\bblam}\hat{\ccalL}_t(\bbx_t,\bblam_t)\|^2 . \nonumber
\end{align}
Reorder terms in the above expression such that the gradient-iterate inner product term is on the left-hand side as
\begin{align} \label{eq:dual_grad_neq}
  \nabla_{\bblam}&\hat{\ccalL}_t(\bbx_t,\bblam_t)^T(\bblam_t-\bblam) \\ 
	&\geq \frac{1}{2\eps} \left( \| \bblam_{t+1}-\bblam\|^2 - \| \bblam_{t}-\bblam\|^2 \right) 
	\!-\!\frac{\eps_t}{2}\| \nabla_{\bblam}\hat{\ccalL}_t(\bbx_t,\bblam_t)\|^2. \nonumber
\end{align}
Note that the online Lagrangian [cf. \eqref{eq:lagrangian}] is a concave function of its Lagrange multipliers, which implies that instantaneous Lagrangian differences for fixed $\bbx_t$ satisfy
\begin{align}  \label{eq:dual_conc}
	 \hat{\ccalL}_t(\bbx_t,\bblam_t) - \hat{\ccalL}_t(\bbx_t,\bblam) 
 	   \geq \nabla_{\bblam}\hat{\ccalL}_t(\bbx_t,\bblam_t)^T(\bblam_t-\bblam).
\end{align} 
By using the lower bound stated in  \eqref{eq:dual_grad_neq} for the right hand side of \eqref{eq:dual_conc}, we may write
\begin{align} \label{eq:lagrange_dual_neq}
 \hat{\ccalL}_t&(\bbx_t,\bblam_t) - \hat{\ccalL}_t(\bbx_t,\bblam) \\  
 	&\geq \frac{1}{2\eps_t} \left( \|\bblam_{t+1}-\bblam\|^2 - \|\bblam_{t}-\bblam\|^2 \right)
	\! -\! \frac{\eps_t}{2}\ \!\| \nabla_{\bblam}\hat{\ccalL}_t(\bbx_t,\bblam_t)\|^2. \nonumber
\end{align}
We now turn to establishing a telescopic property of the instantaneous Lagrangian by combining the expressions in \eqref{eq:lagrange_primal_neq} and \eqref{eq:lagrange_dual_neq}. To do so observe that the term $\hat{\ccalL}_t(\bbx_t,\bblam_t)$ appears in both inequalities. Thus, subtraction in inequality \eqref{eq:lagrange_dual_neq} from those in \eqref{eq:lagrange_primal_neq} followed by reordering terms yields
\begin{align}  \label{eq:primal_dual_combo}
&\hat{\ccalL}_t(\bbx_t,\bblam) - \hat{\ccalL}_t(\bbx,\bblam_t)  \nonumber \\ 
	& \leq \frac{1}{2\eps_t} \big( \|\bbx_t\!-\!\bbx\|^2 \!- \!\|\bbx_{t+1}\!-\!\bbx\|^2 
	  +\|\bblam_{t}-\bblam\|^2 \!-\! \|\bblam_{t+1}\!-\!\bblam\|^2 \big)  \nonumber\\
	&\qquad	 +\frac{\eps}{2} \left(\|\nabla_{\bbx}\hat{\ccalL}_t(\bbx_t,\bblam_t)\|^2 +\| \nabla_{\bblam}\hat{\ccalL}_t(\bbx_t,\bblam_t)\|^2\right), \nonumber
\end{align}
which is as stated in \eqref{eq:lemma1}.
$\qed$

\bibliographystyle{IEEEtran}
\bibliography{bibliography}

   \end{document}